\def\nddLII{\dot\eta_{.2}\,}
\def\nddL{\dot\Omega_{.2}\,}
\def\ndqLII{\dot\eta_{.3}\,}
\def\ndqL{\dot\Omega_{.3}\,}
\def\ndsLII{\dot\eta_{.4}\,}
\def\ndsL{\dot\Omega_{.4}\,}
\def\etdLII{\dot\omega_{.2}\,}
\def\etdL{\dot e_{.2}\,}
\def\etqLII{\dot\omega_{.3}\,}
\def\etqL{\dot e_{.3}\,}
\def\etsLII{\dot\omega_{.4}\,}
\def\etsL{\dot e_{.4}\,}
\def\cross{\bds\times}

\def\nk{n_{\rm b}}

\def\Pb{P_{\rm b}}

\def\rfr#1{Equation~(\ref{#1})}
\def\rfrs#1#2{Equations~(\ref{#1})~to~(\ref{#2})}

\def\derp#1#2{\rp{\partial{#1}}{\partial{#2}}}
\def\dert#1#2{\frac{{{\textrm{d}}}{#1}}{{{\textrm{d}}}{#2}}}

\def\virg#1{``#1"}

\def\eqi{\begin{equation}}
\def\eqf{\end{equation}}
\def\eqia{\begin{eqnarray}}
\def\eqfa{\end{eqnarray}}

\def\rp#1#2{{#1\over#2}}
\def\lb#1{\label{#1}}
\def\kap{\bds{\hat{S}}}

\def\bds#1{\boldsymbol{#1}}

\def\ton#1{\left(#1\right)}
\def\qua#1{\left[#1\right]}
\def\grf#1{\left\{#1\right\}}
\def\ang#1{\left\langle #1\right\rangle}
\documentclass[onecolumn]{aastex}
\usepackage{morefloats}
\usepackage[title]{appendix}
\usepackage{hyperref}
\usepackage{booktabs}
\usepackage[table,xcdraw]{xcolor}
\usepackage{multirow}
\usepackage{rotating,tabularx}
\usepackage{float}
\usepackage{enumerate}
\usepackage{rotating}
\usepackage[polutonikogreek,english]{babel}
\usepackage{amsmath,starfont,textgreek,w-greek,wasysym}
\usepackage{amsthm}
\usepackage{amscd,lineno}
\usepackage{amssymb,dsfont}
\usepackage{graphicx,epsfig}
\usepackage{txfonts}
\usepackage{xr-hyper}

\RequirePackage{color}

\newcommand{\emaila}{lorenzo.iorio@libero.it}

\allowdisplaybreaks[1]

\begin{document}

\title{A HERO for general relativity}

\shortauthors{L. Iorio}

\author{Lorenzo Iorio\altaffilmark{1} }
\affil{Ministero dell'Istruzione, dell'Universit\`{a} e della Ricerca
(M.I.U.R.)-Istruzione
\\ Permanent address for correspondence: Viale Unit\`{a} di Italia 68, 70125, Bari (BA),
Italy}

\email{\emaila}

\begin{abstract}
HERO (Highly Eccentric Relativity Orbiter) is a  space-based mission concept aimed to perform several tests of post-Newtonian gravity around the Earth with a preferably drag-free spacecraft moving along a highly elliptical path fixed in its plane undergoing a relatively fast secular precession. We considered two possible scenarios: a fast, 4-hr orbit with high perigee height of $1,047\,\mathrm{km}$, and a slow, 21-hr path with a low perigee height of $642\,\mathrm{km}$. HERO may detect, for the first time, the post-Newtonian orbital effects induced by the mass quadrupole moment $J_2$ of the Earth which, among other things, affects the semimajor axis $a$ via a secular trend of $\simeq 4-12\,\mathrm{cm\,yr}^{-1}$, depending on the orbital configuration. Recently, the secular decay of the semimajor axis of the passive satellite LARES was measured with an error as little as $0.7\,\mathrm{cm\,yr}^{-1}$. Also the post-Newtonian spin dipole (Lense-Thirring) and mass monopole (Schwarzschild) effects could be tested to a high accuracy depending on the level of compensation of the non-gravitational perturbations, not treated here. Moreover, the large eccentricity of the orbit would allow to constrain several long-range modified models of gravity and to accurately measure the gravitational red-shift as well. Each of the six Keplerian orbital elements could be individually monitored to extract the $GJ_2/c^2$ signature, or they could be suitably combined in order to disentangle the post-Newtonian effect(s) of interest from the competing mismodeled Newtonian secular precessions induced by the zonal harmonic multipoles $J_\ell$ of the geopotential. In the latter case, the systematic uncertainty due to the current formal errors $\upsigma_{J_\ell}$ of a recent global Earth's gravity field model are better than $1\%$ for all the post-Newtonian effects considered, with a peak of $\simeq 10^{-7}$ for the Schwarzschild-like shifts. Instead, the gravitomagnetic spin octupole precessions are too small to be detectable.
\end{abstract}

keywords{
General relativity and gravitation; Experimental studies of gravity; Experimental
tests of gravitational theories; Satellite orbits
}
\section{Introduction}
The (slow) motion of a test particle  moving in the spacetime (weakly) deformed by the mass-energy content of an isolated, axially symmetric rotating body of mass $M$, angular momentum $\bds S$, polar and equatorial radii $R_\textrm{p},~R_\textrm{e}$, ellipticity $\varepsilon = \sqrt{1- R^2_\textrm{p}/R^2_\textrm{e}}$, dimensionless quadrupole mass moment $J_2$ exhibits several  post-Newtonian (pN) features. Some of them have never been put to the test so far because of their smallness; they are the gravitoelectric and gravitomagnetic effects associated with the asphericity of the central body induced by its mass quadrupole and spin octupole moments, respectively \citep{1988CeMec..42...81S, Sof89, 1991ercm.book.....B, 2014PhRvD..89d4043W,2014CQGra..31x5012P,2015CeMDA.123....1M}.

Instead, the pN orbital effects which have been extensively tested so far in several terrestrial and astronomical scenarios are the gravitoelectric and gravitomagnetic precessions due to the mass monopole and spin dipole moments, respectively. The former is responsible of the time-honored, previously anomalous perihelion precession of Mercury \citep{LeVer1859}, whose explanation by \citep{Ein15} was the first empirical success of his newly born theory of gravitation. It was later repeatedly measured with radar measurements of Mercury itself \citep{1972PhRvL..28.1594S,1990grg..conf..313S}, of other inner planets \citep{1978AcAau...5...43A,1992mgm..conf..353A}, and of the asteroid Icarus \citep{1968PhRvL..20.1517S,1971AJ.....76..588S} as well. Also binary pulsars \citep{2006Sci...314...97K} and Earth's artificial satellites \citep{2010PhRvL.105w1103L,2014PhRvD..89h2002L} have been used so far.
The latter is the so-called Lense-Thirring effect \citep{LT18,2007GReGr..39.1735P,2008mgm..conf.2456P,Pfister2012,Pfister2014} which is currently under scrutiny in the Earth's surrounding with the geodetic satellites of the LAGEOS family; see, e.g., \citet{2013CEJPh..11..531R},\ and \citet{Lucchesi019}, and references therein. Another gravitomagnetic effect-the Pugh-Schiff rates of change of orbiting gyroscopes \citep{Pugh59,Schiff60}-was successfully tested in the field of the Earth with the dedicated Gravity Probe B (GP-B) spaceborne mission a few years ago \citep{2011PhRvL.106v1101E,2015CQGra..32v4001E} to a $19\%$ accuracy level, despite the originally expected one was of the order of $1\%$ \citep{2001LNP...562...52E}.

By assuming the validity of general relativity, its mass quadrupole and spin octupole accelerations are, to the first post-Newtonian (pN) order,
\begin{align}
{\bds A}^{\textrm{pN}M J_2} \lb{pNMJ2} &=  \rp{\mu J_2 R_\textrm{e}^2}{c^2 r^4}\grf{\rp{3}{2}\qua{\ton{5\xi^2 - 1}\bds{\hat{r}} - 2\xi\kap  }\ton{{\mathrm{v}}^2 -\rp{4\mu}{r} }- 6\qua{\ton{5\xi^2 - 1}{\mathrm{v}}_r  - 2\xi{\mathrm{v}}_S   }\bds{\mathrm{v}}
- \rp{2\mu}{r}\ton{3\xi^2 - 1}\bds{\hat{r}}},\\ \nonumber \\
\bds A^{\textrm{pN}SJ_2} &= \rp{3GSR_\textrm{e}^2\varepsilon^2 }{7c^2r^5}\bds{\mathrm{v}}\bds\times\grf{5\xi\qua{7\xi^2 - 3}\bds{\hat{r}} + 3\qua{1 - 5\xi^2}\kap},\lb{pNSJ2}
\end{align}
where $G$ is the Newtonian constant of gravitation, $\mu\doteq GM$ is the gravitational parameter of the primary, $c$ is the speed of light in vacuum,
$\bds{\hat{S}}$ is the unit vector of the rotational axis,
\eqi
\xi\doteq \kap\bds\cdot\bds{\hat{r}}
\eqf
is the cosine of the angle between the directions of the body's angular momentum and the orbiter's position vector $\bds r$, and
\begin{align}
{\mathrm{v}}_r &\doteq \bds{\mathrm{v}}\bds\cdot\bds{\hat{r}}, \\ \nonumber \\
{\mathrm{v}}_S &\doteq \bds{\mathrm{v}}\bds\cdot\kap
\end{align}
are the components of the particle's velocity $\bds{\mathrm{v}}$ along the radial direction and the primary's spin respectively.
The averaged rates of change of the semimajor axis $a$, the eccentricity $e$, the inclination $I$, the longitude of the ascending node $\Omega$ and the argument of pericenter $\omega$ induced by \rfrs{pNMJ2}{pNSJ2} were analytically calculated for a general orientation of $\bds{\hat{S}}$ in space by \citet{2015IJMPD..2450067I,2019MNRAS.484.4811I}; previous derivations of the gravitoelectric mass quadrupole effects in the particular case of an equatorial coordinate system with its reference $z$ axis aligned with $\bds{\hat{S}}$ can be found in \citet{1988CeMec..42...81S, Sof89, 1991ercm.book.....B, 2014PhRvD..89d4043W}.

In this paper, we will preliminarily explore the perspectives of measuring, for the first time, some consequences of \rfrs{pNMJ2}{pNSJ2} by suitably designing a dedicated drag-free satellite-based mission around the Earth encompassing a highly eccentric geocentric orbit exploiting the frozen perigee configuration; we provisionally name it as HERO (Highly Eccentric Relativity Orbiter). For some embryonal thoughts about the possibility of using an Earth's spacecraft to measure the pN gravitoelectric effects proportional to $GJ_2/c^2$, see \citet{2013CQGra..30s5011I,2015IJMPD..2450067I}; for deeper investigations concerning a possible probe around Jupiter to measure them and the pN gravitomagnetic signature proportional to $GS\varepsilon^2/c^2$, see \citet{2013CQGra..30s5011I, 2019MNRAS.484.4811I}. About the propagation of the electromagnetic waves in the deformed spacetime of an oblate body and the perspectives of measuring the resulting deflection due to Jupiter with astrometric techniques, see, e.g.,
\citet{2006CQGra..23.4853C,2007PhRvD..75f2002K,2008PhRvD..77d4029L,2019MNRAS.485.1147A}, and references therein.
We will show that the size of the secular rate of $a$ predicted by \rfr{pNMJ2} falls within the recently reached experimental accuracy in measuring phenomenologically such a kind of an effect with the existing passive LARES satellite \citep{Lucchesi019}. Be that as it may, we will show that, as a by-product, also other general relativistic features of motion could be measured with high accuracy, at least as far as the systematic error due to the current formal level of mismodeling in the competing classical precessions due to the zonal harmonic coefficients $J_\ell,\,\ell=2,\,3,\,4,\ldots$ of the multipolar expansion of the Earth's gravity potential is concerned.
To this aim, it is crucial to assess the level of  possible cancelation of the non-gravitational perturbations by the drag-free technology. Its evaluation is outside the scopes of the present paper. However, we will look in detail at the atmospheric drag, which is one of the major disturbing non-conservative accelerations inducing relevant competing signatures, especially on $a$. For the sake of simplicity, we will assume a spherical shape for a passive spacecraft.

The high eccentricity of the suggested orbit of HERO would allow also for accurate tests of the gravitational redshift provided that the spacecraft is endowed with accurate atomic clocks; see Table\,\ref{tavola1} and Table\,\ref{tavola3} for the expected sizes of it. For recent tests of such an effect performed with the H-maser clocks carried onboard the satellites GSAT0201 (5-Doresa) and GSAT0202 (6-Milena) of the Galileo constellation by exploiting their fortuitous rather high eccentricity due to their erroneous orbital injection, see \citet{2018PhRvL.121w1102H,2018PhRvL.121w1101D}.

Finally, several long-range modified models of gravity  \citep{2012PhR...513....1C} imply spherically symmetric modifications of the Newtonian inverse-square law which induce net secular precessions of the pericenter and the mean anomaly at epoch. They would represent further valuable goals for HERO.

The paper is organized as follows. Section\,\ref{confis} is the main body of the paper; it discusses two possible orbital configurations along with the magnitude of the various pN effects and the size of the corresponding systematic errors due to the current level of mismodeling in the multipolar zonal coefficients of the geopotential. It deals also with the linear combination approach which could be implemented in order to reduce the bias due to the latter ones. Section\,\ref{fine} summarizes our findings and offer our conclusions.
Appendix\,\ref{appena} contains the tables and the figures. Appendix\,\ref{appenb} displays the analytically calculated orbital precessions due to the zonal harmonics of the geopotential up to degree $\ell=8$. Appendix\,\ref{appendrag} deals in detail with the impact of the atmospheric drag on all the orbital elements of a spherical, passive geodetic satellite in a highly elliptical orbit.
\section{Two different orbital configurations for HERO}\lb{confis}
In Table\,\ref{tavola1} and Table\,\ref{tavola3}, two different orbital configurations are proposed. They imply highly eccentric orbits, characterized by values of the eccentricity as large as $e=0.45$ and $e=0.82$, respectively, and the critical inclination $I_\mathrm{crit} = \arcsin\ton{2/\sqrt{5}}$ which allows to keep the argument of perigee $\omega$ essentially fixed over any reasonable time span for an actual data analysis and the longitude of the ascending node $\Omega$ circulating at a relatively high pace. Their orbital periods are $\Pb=4.3\,\mathrm{hr}$ and $\Pb=21.3\,\mathrm{hr}$, with perigee heights of $h_\mathrm{min}=1,047\,\mathrm{km}$ and $h_\mathrm{min}=642\,\mathrm{km}$, respectively.

One of the most interesting relativistic features of motion is, perhaps, the relatively large value of the expected semimajor axis increase $\ang{\dot a}$ induced by the pN gravitoelectric quadrupolar acceleration of \rfr{pNMJ2}; let us recall that it is \citep{1988CeMec..42...81S,1991ercm.book.....B,2013CQGra..30s5011I}
\eqi
\ang{\dert a{t}} = \rp{9\,a\,\nk^3\,R_\mathrm{e}^2\,J_2\,e^2\,\ton{6+e^2}\,\sin^2 I\,\sin2\omega}{8\,c^2\,\ton{1-e^2}^{4}},
\eqf where  $\nk=\sqrt{\mu/a^3}=2\uppi/\Pb$ is the Keplerian mean motion. Indeed, according to Table\,\ref{tavola2} and Table\,\ref{tavola4}, its predicted rates for the orbital geometries considered in Table\,\ref{tavola1} and Table\,\ref{tavola3} are $\ang{\dot a} =3.8\,\mathrm{cm\,yr}^{-1}$ and $\ang{\dot a} =11.6\,\mathrm{cm\,yr}^{-1}$, respectively.
Suffice it to say that for the existing passive satellites of the LAGEOS family, whose orbits are essentially circular, secular decay rates have been measured over the last decades with an experimental accuracy of $\upsigma_{\ang{\dot a}}\simeq 3\,\mathrm{cm\,yr}^{-1}$ for LAGEOS \citep{1982CeMec..26..361R,Sosproc,Sosbook}, and $\upsigma_{\ang{\dot a}}=0.7\,\mathrm{cm\,yr}^{-1}$ for LARES \citep{Lucchesi019}. It is arguable that an active mechanism of compensation of the non-gravitational accelerations would allow to increase such accuracies, allowing, perhaps, to measure the pN quadrupolar effect at a $\simeq 1-10\%$ level, depending on the orbital configuration adopted. As far as possible competing effects of gravitational origin are concerned,  neither the static and time-dependent parts of the geopotential nor 3rd-body lunisolar attractions induce nonvanishing averaged perturbations on $a$.
Thus, it is of the utmost importance the reduction of the non-conservative accelerations. Among them, a prominent role is played the atmospheric drag, treated in detail in Appendix\,\ref{appendrag}, because it causes a net long-term averaged decay rate of $a$. By modeling the spacecraft as a LARES-like cannonball geodetic satellite for the sake of simplicity, it can be shown that, in the case of the orbital configuration of Table\,\ref{tavola1}, the average acceleration due to the neutral drag only amounts to $\ang{A}_\mathrm{drag} = (2.3-0.8)\times 10^{-11}\,\mathrm{m\,s}^{-2}$ over one orbital period, so that the resulting effect on $a$ would be as large as $\ang{\dot a}_\mathrm{drag} = -(5.1-2.0)\,\mathrm{m\,yr}^{-1}$; see Table\,\ref{tavola2ter}, and Figures\,\ref{fig1}\,to\,\ref{fig2}. For the more eccentric orbital configuration of Table\,\ref{tavola3}, we have $\ang{A}_\mathrm{drag} = (1.22-7.3)\times 10^{-11}\,\mathrm{m\,s}^{-2}$ and $\ang{\dot a}_\mathrm{drag} = -(27.6-164.6)\,\mathrm{m\,yr}^{-1}$, as shown by Table\,\ref{tavola4ter}, and Figures\,\ref{fig3}\,to\,\ref{fig4}. An inspection of Figure\,\ref{fig1} and Figure\,\ref{fig3} reveals that, as expected, most of the disturbing effect occurs around the perigee passage, i.e. for $f\simeq 0$. This may help in suitably calibrating the counteracting action of the drag-free mechanism.

Table\,\ref{tavola2} and Table\,\ref{tavola3} show that also all the other Keplerian orbital elements exhibit non-zero secular rates of change due to \rfr{pNMJ2}.
This is a potentially important feature since they could be linearly combined, as suggested by\footnote{In that case, the aliasing Newtonian effect which must be disentangled from the pN perihelion precessions is due to the quadrupole mass moment $J_2$ of the Sun.} \citet{1990grg..conf..313S} in a different context, in order to decouple the pN effect(s) of interest from the disturbing mismodeled Newtonian secular precessions induced by the Earth's zonal multipoles. Indeed, contrary to $a$, the other Keplerian orbital elements do exhibit long-term averaged precessions due to the classical zonal harmonics $J_\ell,\,\ell=2,\,3,\,4,\ldots$ of the geopotential; they are analytically calculated in Appendix\,\ref{appenb} up to degree $\ell=8$. Depending on the specific orbital geometry, they can be both secular and harmonic, or entirely\footnote{Indeed, the harmonic perturbations contain also the satellite's perigee which, for the critical inclination $I_\mathrm{crit}=\arcsin\ton{2/\sqrt{5}}$ adopted in Table\,\ref{tavola1} and Table\,\ref{tavola3}, stays essentially constant becoming \virg{frozen}. In this case, also the long-term, harmonic rates become secular trends.} secular.
To this aim, we complete the set of the pN orbital effects by analytically calculating the averaged rate of the mean anomaly at epoch due to the pN accelerations considered. It turns out that \rfrs{pNMJ2}{pNSJ2} induce
\begin{align}
\ang{\dot\eta} &=\rp{\mu\,\nk\,R^2\,J_2\,\qua{-\ton{80 + 73\,e^2}\,\ton{1 + 3\,\cos 2I} - 84\,\ton{1 + 2\,e^2}\,\sin^2 I\,\cos 2 \omega}}
{32\,c^2\,a^3\,\ton{1 - e^2}^{5/2}}, \\ \nonumber \\
\ang{\dot\eta} &= \rp{9\,G\,S\,R^2\,\varepsilon^2\,\qua{5\,\cos 3 I + \cos I\,\ton{3 + 10\,\sin^2 I\,\cos 2\omega}}}{56\,a^5\,c^2\,\ton{1 - e^2}^2},
\end{align}
respectively, while the gravitoelectric mass monopole acceleration yields
\eqi
\ang{\dot\eta} = \rp{\mu\,\nk\,\qua{-15 + 6\,\sqrt{1 - e^2}  + \ton{9 - 7\,\sqrt{1 - e^2}}\,\zeta}}{c^2\,a\,\sqrt{1 - e^2}},
\eqf
where $\zeta\doteq M\,m\,/\,\ton{M+m}^2$, and $m$ is the satellite's mass. Instead, it turns out that there is no net Lense-Thirring effect on $\eta$.
To the benefit of the reader, we review the linear combination approach, which is a generalization of that proposed explicitly for the first time by \citet{1996NCimA.109..575C} to test the pN Lense-Thirring effect in the gravitomagnetic field of the Earth with the artificial satellites of the LAGEOS family. It should be noted that, actually, it is quite general, being not necessarily limited just to the pN spin dipole case.
By looking at $N$ orbital elements\footnote{At least one of them must be affected by the pN effect one is looking for. In principle, the $N$ orbital elements $\kappa^{(i)}$ may be different from one another belonging to the same satellite, or some of them may be identical belonging to different spacecraft (e.g., the nodes of two different vehicles).}
$\kappa^{(i)}$  experiencing classical long-term precessions due to the zonals of the geopotential, the following $N$ linear combinations can be written down
\eqi
\upmu_\mathrm{pN}\,\ang{\dot\kappa}_\mathrm{pN}^{(i)}+ \sum_{s=2}^{N}\,\ton{\derp{\ang{\dot\kappa}_{J_{s}}^{(i)}}{J_{s}}}\,\delta J_{s},\,i=1,2,\ldots N. \lb{lin}
\eqf
They involve the pN averaged precessions $\ang{\dot\kappa}_\mathrm{pN}^{(i)}$ as predicted by general relativity and scaled by a multiplicative parameter $\upmu_\mathrm{pN}$, and the errors in the computed  secular node precessions due to the uncertainties  in the first $N-1$ zonals $J_{s},~s=2,3,\ldots N$, assumed as mismodeled through $\delta{J_{s}},~s=2,3,\ldots N$.
In the following, we will use the shorthand
\eqi
\dot\kappa_{.\ell}\doteq\derp{\ang{\dot\kappa}_{J_\ell}}{J_\ell}
\eqf
for the partial derivative of the classical averaged precession $\ang{\dot\kappa}_{J_\ell}$ with respect to the generic even zonal $J_\ell$ of degree $\ell$; see Appendix\,\ref{appenb}. Then, the $N$ combinations of \rfr{lin} are posed equal to the experimental residuals $\delta\dot\kappa^{(i)},~i=1,2,\ldots N$ of each of the $N$ orbital elements considered. In principle, such residuals account for the purposely unmodelled pN effect, the mismodelling of the static and time-varying parts of the geopotential, and the non-gravitational forces. Thus, one gets
\eqi
\delta\dot\kappa^{(i)} = \upmu_\mathrm{pN}\,\ang{\dot\kappa}_\mathrm{pN}^{(i)}+ \sum_{s=2}^{N}\,\dot\kappa^{(i)}_{.s}\,\delta J_{s},~i=1,2,\ldots N. \lb{lineq} \eqf
If we look at the pN scaling parameter\footnote{In general, it is not necessarily one of the parameters of the parameterized post-Newtonian (PPN) formalism, being possibly a combination of some of them.} $\upmu_\mathrm{pN}$ and the mismodeling in the first $N-1$ zonals $\delta{J_{s}},~s=2,3,\ldots N$ as unknowns, we can interpret \rfr{lineq} as an inhomogenous linear system of $N$ algebraic equations in the $N$ unknowns
\eqi
\underbrace{\upmu_\mathrm{pN},~\delta J_2,~\delta J_3 \ldots \delta J_{N}}_{N},
\eqf
whose coefficients are
\eqi
\ang{\dot\kappa}^{(i)}_\mathrm{pN},\,\dot\kappa^{(i)}_{.s},\,i=1,2,\ldots N,\,s=2,3,\ldots N,
\eqf
while the constant terms are the $N$ orbital residuals
\eqi
\delta\dot\kappa^{(i)},\,i=1,2,\ldots N.
\eqf
It turns out that, after some algebraic manipulations,  the dimensionless pN scaling parameter, which is 1 in general relativity, can be expressed as
\eqi
\upmu_\mathrm{pN}=\rp{\mathcal{C}_\delta}{\mathcal{C}_\mathrm{pN}}.\lb{rs}
\eqf
In \rfr{rs}, the combination of the $N$ orbital residuals
\eqi
{\mathcal{C}}_\delta \doteq \delta\dot\kappa^{(1)} + \sum_{j=1}^{N-1}\,c_j\,\delta\dot\kappa^{(j+1)} \lb{combo}
\eqf
is, by construction, independent of the first $N-1$ zonals, being impacted by the other ones of degree $\ell > N$ along with the non-gravitational perturbations and other possible orbital perturbations which cannot be reduced to the same formal expressions of the first $N-1$ zonal rates. Instead,
\eqi
\mathcal{C}_\mathrm{pN} \doteq \ang{\dot\kappa}_\mathrm{pN}^{(1)} + \sum_{j=1}^{N-1}\,c_j\,\ang{\dot\kappa}_\mathrm{pN}^{(j+1)} \lb{combopN}\eqf
combines the $N$ pN orbital precessions as predicted by general relativity.
The dimensionless coefficients $c_j,\ j=1,2,\ldots N-1$ in \rfr{combo}-\rfr{combopN} depend only on some of the orbital parameters of the satellite(s) involved in such a way that, by construction, ${\mathcal{C}}_\delta=0$ if \rfr{combo} is calculated by posing
\eqi
\delta\dot\kappa^{(i)}=\dot\kappa^{(i)}_{.\ell}\,\delta J_{\ell},\ i=1,2,\ldots N
\eqf
for any of the first $N-1$ zonals, independently of the value assumed for its uncertainty $\delta{J_{\ell}}$.

As far as HERO is concerned, the linear combination of the four experimental residuals $\delta\Omega,\,\delta\eta,\,\delta e,\,\delta\omega$ of the satellite's node, mean anomaly at epoch, eccentricity and perigee suitably designed to cancel out the secular precessions due to the first three zonal harmonics $J_2,\,J_3,\,J_4$ of the geopotential is
\eqi
\mathcal{C}_\delta=\delta\Omega + c_1\,\delta\eta + c_2\,\delta e + c_3\,\delta\omega.\lb{kombo}
\eqf
The coefficients $c_1,~c_2,~c_3$  turn out to be
\begin{align}
c_1 \lb{c1}& = \rp{\nddL\etqL\etsLII -\etdL\ndqL\etsLII -\nddL\etqLII\etsL +\etdLII\ndqL\etsL +\etdL\etqLII\ndsL -\etdLII\etqL\ndsL}{\etdL\ndqLII\etsLII -\nddLII\etqL\etsLII -\etdL\etqLII\ndsLII +\etdLII\etqL\ndsLII +\nddLII\etqLII\etsL -\etdLII\ndqLII\etsL},\\\nonumber\\
c_2 \lb{c2} &= \rp{-\nddL\ndqLII\etsLII +\nddLII\ndqL\etsLII +\nddL\etqLII\ndsLII -\etdLII\ndqL\ndsLII -\nddLII\etqLII\ndsL +\etdLII\ndqLII\ndsL}{\etdL\ndqLII\etsLII -\nddLII\etqL\etsLII -\etdL\etqLII\ndsLII +\etdLII\etqL\ndsLII +\nddLII\etqLII\etsL -\etdLII\ndqLII\etsL},\\\nonumber\\
c_3 \lb{c3} &= \rp{-\nddL\etqL\ndsLII +\etdL\ndqL\ndsLII +\nddL\ndqLII\etsL -\nddLII\ndqL\etsL -\etdL\ndqLII\ndsL +\nddLII\etqL\ndsL}{\etdL\ndqLII\etsLII -\nddLII\etqL\etsLII -\etdL\etqLII\ndsLII +\etdLII\etqL\ndsLII +\nddLII\etqLII\etsL -\etdLII\ndqLII\etsL}.
\end{align}
Their numerical values, computed with the formulas of Appendix\,\ref{appenb} for the orbital configurations of Table\,\ref{tavola1} and Table\,\ref{tavola3}, are listed in Table\,\ref{tavola2bis} and Table\,\ref{tavola4bis}, respectively. In them, the combined mismodeled classical precessions due to the uncancelled zonals, calculated by assuming the formal, statistical sigmas $\upsigma_{J_\ell},\,\ell=5,\,6,\ldots$ of the recent global gravity field solution Tongji-Grace02s \citep{2018JGRB..123.6111C} as a measure of their uncertainties $\delta J_\ell,\,\ell=5,\,6,\ldots$, are reported as well. However, caution is in order since the realistic level of mismodeling in the geopotential's coefficients is usually larger than the mere formal errors released in the models produced by various institutions and publicly available on the Internet at http://icgem.gfz-potsdam.de/tom$\_$longtime. A correct evaluation of the actual uncertainties in the zonal harmonics require great care by suitably comparing different global gravity field models; we will not deal with such a task here.
%
From an inspection of Table\,\ref{tavola2bis} and Table\,\ref{tavola4bis}, it can be noted that the (formal) impact of the uncancelled zonals on the combined pN mass quadrupole effect ($GJ_2/c^2$) is at the $\simeq 0.6-0.07\%$ level for the proposed orbital configurations of Table\,\ref{tavola1} and Table\,\ref{tavola3}. If the pN spin dipole Lense-Thirring effect ($GS/c^2$) is considered, the systematic error due to the mismodeling in $J_\ell,\,\ell>4$ is about $\simeq 0.1-0.03\%$. The pN mass monopole combined precessions ($GM/c^2$) are affected at the $\simeq (20-5)\times 10^{-7}$ relative level. Instead, it turns out that the the combined mismodelled classical precessions are at the same level of the pN spin octupole trends ($GS\varepsilon^2/c^2$). It may  not be unrealistic  to expect that, when the forthcoming global gravity field models based on the analysis of the entire long data records of the dedicated GRACE and GOCE missions will be finally available, the current merely formal level of uncertainties in the geopotential's zonal harmonics may be considered as realistic.
Moreover, in the next years, the mission GRACE-FO (GFO) \citep{2012JGeod..86.1083S}, launched in May 2018, will also contribute to the production of new global Earth's gravity field models of increased quality.

On the other hand, the size of the coefficients $c_1,\,c_2,\,c_3$ amplifies the impact of any non-gravitational perturbations that may affect the spacecraft; thus, they should be effectively counteracted  by some active drag-free apparatus. In particular, the coefficient $c_2$ of the eccentricity is $\simeq 30-40$; Table\,\ref{tavola2ter} and Table\,\ref{tavola4ter} show that the expected secular decrease rate of $e$ due to the atmospheric drag is rather large. Thus, some trade-off may be required among the need of reducing the systematic error of gravitational origin and the actual performance of the drag-free mechanism by looking, e.g., at different linear combinations.
It may be interesting to note the case of the mean anomaly at epoch. Indeed, in the case of the high perigee orbital configuration of Table\,\ref{tavola1}, Table\,\ref{tavola2} and Table\,\ref{tavola2ter} tell us that the neutral atmospheric drag would represent just $\simeq 1-2\%$ of the predicted pN $GJ_2/c^2$ precession on $\eta$. On the other hand, the present-day formal mismodeling in the classical $J_2$-induced rate is about $19\%$ of it. If the pN Schwarzschild-like effect is considered, the formal bias due to  $J_2$ is at the $\simeq 10^{-5}$ level, while the impact of the atmospheric drag is as little as $\simeq 10^{-6}$. The neutral atmospheric drag has a larger impact on the pN precessions of $\eta$ in the case of the low perigee configuration of Table\,\ref{tavola3}, as shown by Table\,\ref{tavola4} and Table\,\ref{tavola4ter}.
\section{Summary and overview}\lb{fine}
The HERO concept, meant as a hopefully drag-free spacecraft moving in a highly eccentric orbit in a frozen perigee configuration aimed to perform several tests of relativistic gravity in the Earth's spacetime, represents, in principle, a promising opportunity to measure, for the first time, a general relativistic effect which has never received the same attention of the more known Schwarzschild and Lense-Thirring precessions so far: the post-Newtonian gravitoelectric orbital shifts due to the mass quadrupole moment of the Earth. Indeed, the systematic uncertainty in the combined satellite's precessions due to the formal, statistical errors in the competing Newtonian mass multipoles of the geopotential, as per one of the most recent global gravity field models, is currently below the per cent level for both the orbital configurations proposed.
A unique feature of such a post-Newtonian effect is that also the semimajor axis $a$ undergoes a long-term variation which, for a frozen perigee configuration, resembles a secular trend of the order of $\simeq 4-11\,\mathrm{cm\,yr}^{-1}$, depending on the orbital geometry chosen. At present, the secular decay of the semimajor axis of the existing passive geodetic satellite LARES has been measured to an accuracy better than $1\,\mathrm{cm\,yr}^{-1}$ at $2\upsigma$ level.
As far as the traditional Lense-Thirring and Schwarzschild-like post-Newtonian precessions, the formal systematic bias due to the present-day mismodeling in the classical Earth's zonal harmonics is currently $\lesssim 0.1\%$ and $\lesssim 0.0002\%$, respectively if a suitable linear combination of some of the orbital elements of HERO is adopted. However, it must be stressed that the actual uncertainties in the zonal multipoles of the terrestrial gravity field may usually be (much) worse than the sigmas released in the various global gravity solutions. Nonetheless, it cannot be ruled out that, if and when HERO will fly, our knowledge of the Earth's gravity field will have reached such levels that today's only formal uncertainties can finally be considered as truly realistic.
In addition to the post-Newtonian accelerations, HERO may perform an accurate test of the gravitational red-shift in view of its high eccentricity. Also several models of modified gravity, which generally affect the perigee and the mean anomaly at epoch with secular precessions, could be fruitfully put to the test. A crucial aspect is represented by the level of compensation of the non-gravitational perturbation which will be practically attainable with some drag-free apparatus; suffice it to say that the nominal size of the competing secular decrease of the semimajor axis due to, e.g., the neutral atmospheric drag can reach the $\simeq 5-160\,\mathrm{m\,yr}^{-1}$ level if a passive, cannonball satellite is considered. Its investigation deserves a dedicated publication.

\begin{appendices}
\section{Tables and figures}\lb{appena}
\renewcommand{\thetable}{A\arabic{table}}
\setcounter{table}{0}
\renewcommand{\thefigure}{A\arabic{figure}}
\setcounter{figure}{0}
\begin{table}
\caption{Relevant physical parameters of the Earth \citep{1981CeMec..25..169S,2010ITN....36....1P,2009P&SS...57.1405D}. The zonal harmonics of the geopotential of degree $\ell$ are given by $J_\ell = -\sqrt{2\ell +1}\,{\overline{C}}_{\ell,0},~\ell=2,3,4,\ldots$, where ${\overline{C}}_{\ell,0},~\ell=2,3,4,\ldots$ are the fully normalized Stokes coefficients of degree $\ell$ and order $\textrm{m}=0$ of the multipolar expansion of the Newtonian part of the Earth's gravity field.  The formal, statistical errors in the first seven Stokes coefficients of the geopotential, along with their nominal values, were retrieved from the global gravity field solution Tongji-Grace02s \citep{2018JGRB..123.6111C} retrievable on the Internet at http://icgem.gfz-potsdam.de/tom$\_$longtime.}\lb{tavola0}
\begin{center}
\begin{tabular}{|l|l|l|}
\hline
Physical parameter & Numerical value & Units\\
\hline
Newtonian constant of gravitation $G$  & $6.67259\times 10^{-11}$ & $\textrm{kg}^{-1}~\textrm{m}^3~\textrm{s}^{-2}$ \\
Speed of light in vacuum  $c$ & $2.99792458\times 10^8$ & $\textrm{m}~\textrm{s}^{-1}$\\
%
%
Gravitational parameter $\mu$  & $3.986004418\times 10^{14}$ & $\textrm{m}^3~\textrm{s}^{-2}$ \\
Angular speed $\Psi$ & $7.29\times 10^{-5}$ & $\textrm{s}^{-1}$ \\
Equatorial radius $R_\textrm{e}$ & $6,378.1370$ & $\textrm{km}$ \\
Polar radius $R_\textrm{p}$ & $6,356.7523$ & $\textrm{km}$ \\
Angular momentum $S$ & $5.86\times 10^{33}$ & $\textrm{J\,s}$ \\
%
%
%
%
Normalized Stokes coefficient  ${\overline{C}}_{2,0}$ & $-4.84165299806\times 10^{-4}$ & - \\
Normalized Stokes coefficient ${\overline{C}}_{3,0}$ & $9.571989759740\times 10^{-7}$ & - \\
Normalized Stokes coefficient ${\overline{C}}_{4,0}$ & $5.399893295930\times 10^{-7}$ & - \\
Normalized Stokes coefficient ${\overline{C}}_{5,0}$ & $6.86499810446677\times 10^{-8}$ & - \\
Normalized Stokes coefficient ${\overline{C}}_{6,0}$ & $-1.49976729587105\times 10^{-7}$ & - \\
Normalized Stokes coefficient ${\overline{C}}_{7,0}$ & $9.05017773295824\times 10^{-8}$ & - \\
Normalized Stokes coefficient ${\overline{C}}_{8,0}$ & $4.94794369681244\times 10^{-8}$ & - \\
Formal error   $\upsigma_{{\overline{C}}_{2,0}}$ & $2.98340899705584\times 10^{-13}   $ & -\\
Formal error   $\upsigma_{{\overline{C}}_{3,0}}$ & $8.39284383652709\times 10^{-14}   $ & -\\
Formal error   $\upsigma_{{\overline{C}}_{4,0}}$ & $4.07426781903578\times 10^{-14}   $ & -\\
Formal error   $\upsigma_{{\overline{C}}_{5,0}}$ & $2.57688174349872\times 10^{-14}   $ & -\\
Formal error   $\upsigma_{{\overline{C}}_{6,0}}$ & $1.89009491873398\times 10^{-14}   $ & -\\
Formal error   $\upsigma_{{\overline{C}}_{7,0}}$ & $1.50081719867797\times 10^{-14}   $ & -\\
Formal error   $\upsigma_{{\overline{C}}_{8,0}}$ & $1.27528335995664\times 10^{-14}   $ & -\\
\hline
\end{tabular}
\end{center}
\end{table}
\begin{table}
\caption{Orbital configuration of the proposed satellite HERO: high perigee case. The orbital motion is rather fast since the orbital period $\Pb$ is as short as $\simeq 4\,\mathrm{hr}$. The period $P_{\omega}$ of the perigee is mainly determined by its secular precession due to $J_3,~J_4$ because of the critical inclination which makes the secular precession due to $J_2$ nominally vanishing. Note the relatively short period $P_{\Omega}$ of the node, amounting to less than $2\,\mathrm{yr}$.
}\lb{tavola1}
\begin{center}
\begin{tabular}{|l|l|l|}
\hline
Orbital and  physical parameter & Numerical value & Units\\
\hline
%
%
%
%
%
%
%
Semimajor axis $a$ & $13,500$ & \textrm{km}\\
Orbital period $P_\textrm{b}$ & $4.33$ & \textrm{hr}\\
Orbital eccentricity $e$ & $0.45$ & - \\
Perigee height $h_\textrm{min}$  & $1,046.86$ & \textrm{km}\\
Apogee height $h_\textrm{max}$  & $13,196.9$ & \textrm{km}\\
Orbital inclination $I$ & $63.43$ & \textrm{deg}\\
Argument of perigee $\omega$ & $45$ & \textrm{deg}\\
Period of the node $P_{\Omega}$ & $-1.94$ & \textrm{yr}\\
Period of the perigee $P_{\omega}$ & $-1,363.4$ & \textrm{yr}\\
Gravitational redshift $\rp{\Delta U}{c^2}$ & $3.7\times 10^{-10}$ & $-$ \\
%
%
%
%
%
%
%
\hline
\end{tabular}
\end{center}
\end{table}
\begin{table}
\caption{Nominal pN (first four rows from the top) and mismodeled Newtonian (first seven rows from the bottom) rates of change, averaged over one orbital revolution, of  the semimajor axis $a$, the eccentricity $e$, the inclination $I$, the longitude of the ascending node $\Omega$, the argument of pericenter $\omega$, and the mean anomaly at epoch $\eta$ for the ideal (no orbital injection error on $I$ assumed) orbital configuration of Table~\ref{tavola1}. The units are $\textrm{cm}~\textrm{yr}^{-1}$ for $\ang{\dot a}$, and $\textrm{mas}~\textrm{yr}^{-1}$ for $\ang{\dot e},\,\ang{\dot I},\,\ang{\dot\Omega},\,\ang{\dot\omega},\,\ang{\dot\eta}$.
The uncertainties in the classical rates of change due to the geopotential are the formal, statistical errors $\upsigma_{J_\ell}$ in  $J_\ell,\,\ell=2,\,3,\,\ldots 8$ of the model Tongji-Grace02s \citep{2018JGRB..123.6111C} quoted in Table\,\ref{tavola0}.
}\lb{tavola2}
\begin{center}
\small{
\begin{tabular}{|l|l|l|l|l|l|l|}
\hline
  & $\ang{\dot a}\,\left(\textrm{cm}\,\textrm{yr}^{-1}\right)$
  & $\ang{\dot e}\,\left(\textrm{mas}\,\textrm{yr}^{-1}\right)$
  & $\ang{\dot I}\,\left(\textrm{mas}\,\textrm{yr}^{-1}\right)$
  & $\ang{\dot\Omega}\,\left(\textrm{mas}\,\textrm{yr}^{-1}\right)$
  & $\ang{\dot\omega}\,\left(\textrm{mas}\,\textrm{yr}^{-1}\right)$
  & $\ang{\dot\eta}\,\left(\textrm{mas}\,\textrm{yr}^{-1}\right)$ \\
\hline
$J_2c^{-2}$ & $3.8$ & $0.42$ & $0.02$ & $0.82$ & $-0.14$ & $0.87$\\
$\varepsilon^2c^{-2}$ & $0$ & $-0.008$ & $0.002$ & $0$ & $0.074$ & $-0.015$\\
$Sc^{-2}$ & $0$ & $0$ & $0$ & $32.323$ & $-43.366$ & $0$\\
$Mc^{-2}$ & $0$ & $0$ & $0$ & $0$ & $3,237.8$ & $-9,292.96$\\
$\upsigma_{J_2}$  & $0$ & $0$ & $0$ & $0.411$ & $0$ & $0.164$\\
$\upsigma_{J_3}$  & $0$ & $0$ & $0$ & $0.057$ & $0.026$ & $0$\\
$\upsigma_{J_4}$  & $0$ & $0.002$   & $0.0006$ & $0.034$ & $0.049$ & $0.004$\\
$\upsigma_{J_5}$  & $0$ & $0.005$   & $0.001$ & $0.010$ & $0.036$ & $0.004$\\
$\upsigma_{J_6}$  & $0$ & $0.003$   & $0.0009$ & $0.002$ & $0.025$ & $0.002$\\
$\upsigma_{J_7}$  & $0$ & $0.002$   & $0.0007$ & $0.002$ & $0.015$ & $0.002$\\
$\upsigma_{J_8}$  & $0$ & $0.001$   & $0.0004$ & $0.004$ & $0.006$ & $0.001$\\
\hline
\end{tabular}
}
\end{center}
\end{table}
\begin{table}
\caption{
Upper three rows: Numerical values of the coefficients $c_1,\,c_2,\,c_3$ of the linear combination of \rfr{kombo} canceling out the classical precessions induced by the first three zonal harmonics $J_2,\,J_3,\,J_4$ for the orbital configuration of Table\,\ref{tavola1}.
Middle rows: Uncancelled mismodeled precessions due to the zonal harmonics $J_5,\,J_6,\,J_7,\,J_8$, in $\mathrm{mas\,yr}^{-1}$, linearly combined according to \rfr{kombo}; the formal, statistical errors $\upsigma_{J_\ell}$ in  $J_\ell,\,\ell=5,\,6,\,7,\,8$ of the model Tongji-Grace02s \citep{2018JGRB..123.6111C}, quoted in Table\,\ref{tavola0}
were used.
Lower four rows: pN precessions, in $\mathrm{mas\,yr}^{-1}$, linearly combined according to \rfr{kombo}.
}\lb{tavola2bis}
\begin{center}
\small{
\begin{tabular}{|l|l|l|}
\hline
$c_1$ & $-2.51065$ & $-$\\
$c_2$ & $29.0889$ & $-$\\
$c_3$ & $2.13813$ & $-$\\
\hline
$\upsigma_{J_5}\,\mathrm{formal}$ & $0.06$ & $\mathrm{mas\,yr}^{-1}$\\
$\upsigma_{J_6}\,\mathrm{formal}$ & $0.03$ & $\mathrm{mas\,yr}^{-1}$\\
$\upsigma_{J_7}\,\mathrm{formal}$ & $0.03$ & $\mathrm{mas\,yr}^{-1}$\\
$\upsigma_{J_8}\,\mathrm{formal}$ & $0.02$ & $\mathrm{mas\,yr}^{-1}$\\
%
%
%
%
%
\hline
$J_2\,c^{-2}$ & $10.75$ & $\mathrm{mas\,yr}^{-1}$\\
$\varepsilon\,c^{-2}$ & $-0.03$ & $\mathrm{mas\,yr}^{-1}$\\
$S\,c^{-2}$ & $-60.07$ & $\mathrm{mas\,yr}^{-1}$\\
$M\,c^{-2}$ & $30,254.2$ & $\mathrm{mas\,yr}^{-1}$\\
\hline
\end{tabular}
}
\end{center}
\end{table}
\begin{table}
\caption{Numerically integrated nominal rates of change, averaged over one orbital revolution, of  the semimajor axis $a$, the eccentricity $e$, the inclination $I$, the longitude of the ascending node $\Omega$, the argument of pericenter $\omega$, and the mean anomaly at epoch $\eta$ induced by the neutral atmospheric drag for the orbital configuration of Table~\ref{tavola1}. The units are $\textrm{m}~\textrm{yr}^{-1}$ for $\ang{\dot a}$, and $\textrm{mas}~\textrm{yr}^{-1}$ for $\ang{\dot e},\,\ang{\dot I},\,\ang{\dot\Omega},\,\ang{\dot\omega},\,\ang{\dot\eta}$. For the satellite, assumed spherical in shape and passive, we adopted $C_\mathrm{D}=3.5,\,\Sigma=2.69\times 10^{-4}$ as for the existing LARES \citep{2017AcAau.140..469P}. In regard to the Earth's neutral atmospheric density, we adopted $r_0=r_\mathrm{min}=a\ton{1-e}=1,046.86\,\mathrm{km},\,\rho_0=\rho_\mathrm{max}=\ton{7.3-2.8}\times 10^{-15}\,\mathrm{kg\,m}^{-3},\,\rho_\mathrm{min}=0.001\,\rho_\mathrm{L},\,\lambda=872.87\,\mathrm{km}-938.49\,\mathrm{km}$; the neutral atmospheric density at the  height of LAGEOS is $\rho_\mathrm{L}=6.579\times 10^{-18}\,\mathrm{kg\,m}^{-3}$ \citep{2015CQGra..32o5012L}. See Appendix\,\ref{appendrag} for details. Neither approximations in $e$ nor in $\nu\doteq \Psi/\nk$ were used. The value of $\rho_0=\rho_\mathrm{max}$ was kept fixed over one orbital revolution. Cfr. with the analytical plots in Figure\,\ref{fig1} and the numerically produced time series in Figure\,\ref{fig2}.
}\lb{tavola2ter}
\begin{center}
\small{
\begin{tabular}{|l|l|l|l|l|l|l|}
\hline
    $\rho_0\,\ton{\mathrm{kg\,m}^{-3}}$
  & $\ang{\dot a}\,\left(\textrm{m}\,\textrm{yr}^{-1}\right)$
  & $\ang{\dot e}\,\left(\textrm{mas}\,\textrm{yr}^{-1}\right)$
  & $\ang{\dot I}\,\left(\textrm{mas}\,\textrm{yr}^{-1}\right)$
  & $\ang{\dot\Omega}\,\left(\textrm{mas}\,\textrm{yr}^{-1}\right)$
  & $\ang{\dot\omega}\,\left(\textrm{mas}\,\textrm{yr}^{-1}\right)$
  & $\ang{\dot\eta}\,\left(\textrm{mas}\,\textrm{yr}^{-1}\right)$ \\
\hline
$7.3\times 10^{-15}$ & $-5.1$ & $-41$ & $-0.51$ & $-0.21$ & $0.12$ & $-0.02$\\
$2.8\times 10^{-15}$ & $-2$ & $-16$ & $-0.2$ & $-0.07$ & $0.04$ & $-0.01$\\
\hline
\end{tabular}
}
\end{center}
\end{table}
\begin{table}
\caption{Orbital configuration of the proposed satellite HERO: low perigee case. The orbital motion is relatively slow since the orbital period $\Pb$ is as long as more than $21\,\mathrm{hr}$. The period $P_{\omega}$ of the perigee is mainly determined by its secular precession due to $J_3,~J_4$ because of the critical inclination which makes the secular precession due to $J_2$ nominally vanishing. The period $P_{\Omega}$ of the node is rather long, amounting to more than $13\,\mathrm{yr}$.
}\lb{tavola3}
\begin{center}
\begin{tabular}{|l|l|l|}
\hline
Orbital and  physical parameter & Numerical value & Units\\
\hline
%
%
%
%
%
%
%
Semimajor axis $a$ & $39,000$ & \textrm{km}\\
Orbital period $P_\textrm{b}$ & $21.29$ & \textrm{hr}\\
Orbital eccentricity $e$ & $0.82$ & - \\
Perigee height $h_\textrm{min}$  & $641.86$ & \textrm{km}\\
Apogee height $h_\textrm{max}$  & $64,601.9$ & \textrm{km}\\
Orbital inclination $I$ & $63.43$ & \textrm{deg}\\
Argument of perigee $\omega$ & $45$ & \textrm{deg}\\
Period of the node $P_{\Omega}$ & $-13.45$ & \textrm{yr}\\
Period of the perigee $P_{\omega}$ & $-8,186.71$ & \textrm{yr}\\
Gravitational redshift $\rp{\Delta U}{c^2}$ & $5.7\times 10^{-10}$ & $-$ \\
%
%
%
%
%
%
\hline
\end{tabular}
\end{center}
\end{table}
\begin{table}
\caption{Nominal pN (first four rows from the top) and mismodeled Newtonian (first seven rows from the bottom) rates of change, averaged over one orbital revolution, of  the semimajor axis $a$, the eccentricity $e$, the inclination $I$, the longitude of the ascending node $\Omega$, the argument of pericenter $\omega$, and the mean anomaly at epoch $\eta$ for the ideal (no orbital injection error on $I$ assumed) orbital configuration of Table~\ref{tavola3}. The units are $\textrm{cm}~\textrm{yr}^{-1}$ for $\ang{\dot a}$, and $\textrm{mas}~\textrm{yr}^{-1}$ for $\ang{\dot e},\,\ang{\dot I},\,\ang{\dot\Omega},\,\ang{\dot\omega},\,\ang{\dot\eta}$.
The uncertainties in the classical rates of change due to the geopotential are the formal, statistical errors $\upsigma_{J_\ell}$ in  $J_\ell,\,\ell=2,\,3,\,\ldots 8$ of the model Tongji-Grace02s \citep{2018JGRB..123.6111C} quoted in Table\,\ref{tavola0}.
}\lb{tavola4}
\begin{center}
\small{
\begin{tabular}{|l|l|l|l|l|l|l|}
\hline
  & $\ang{\dot a}\,\left(\textrm{cm}\,\textrm{yr}^{-1}\right)$
  & $\ang{\dot e}\,\left(\textrm{mas}\,\textrm{yr}^{-1}\right)$
  & $\ang{\dot I}\,\left(\textrm{mas}\,\textrm{yr}^{-1}\right)$
  & $\ang{\dot\Omega}\,\left(\textrm{mas}\,\textrm{yr}^{-1}\right)$
  & $\ang{\dot\omega}\,\left(\textrm{mas}\,\textrm{yr}^{-1}\right)$
  & $\ang{\dot\eta}\,\left(\textrm{mas}\,\textrm{yr}^{-1}\right)$ \\
\hline
$J_2c^{-2}$ & $11.6$ & $0.115$ & $0.010$ & $0.100$ & $-0.022$ & $0.092$\\
$\varepsilon^2c^{-2}$ & $0$ & $-0.0006$ & $0.0008$ & $0$ & $0.0106$ & $-0.0004$\\
$Sc^{-2}$ & $0$ & $0$ & $0$ & $5.09$ & $-6.83$ & $0$\\
$Mc^{-2}$ & $0$ & $0$ & $0$ & $0$ & $555.661$ & $-1,226.13$\\
$\upsigma_{J_2}$  & $0$ & $0$ & $0$ & $0.059$ & $0$ & $0.015$\\
$\upsigma_{J_3}$  & $0$ & $0$ & $0$ & $0.0128$ & $0.006$ & $0$\\
$\upsigma_{J_4}$  & $0$ & $0.0001$   & $0.0002$ & $0.005$ & $0.007$ & $0.0009$\\
$\upsigma_{J_5}$  & $0$ & $0.0002$   & $0.0003$ & $0.002$ & $0.005$ & $0.0006$\\
$\upsigma_{J_6}$  & $0$ & $0.0002$   & $0.0002$ & $0.0002$ & $0.003$ & $0.0003$\\
$\upsigma_{J_7}$  & $0$ & $0.0001$   & $0.0002$ & $0.0005$ & $0.002$ & $0.0002$\\
$\upsigma_{J_8}$  & $0$ & $0.00008$   & $0.0001$ & $0.0008$ & $0.0007$ & $0.00007$\\
\hline
\end{tabular}
}
\end{center}
\end{table}
\begin{table}
\caption{
Upper three rows: Numerical values of the coefficients $c_1,\,c_2,\,c_3$ of the linear combination of \rfr{kombo} canceling out the classical precessions induced by the first three zonal harmonics $J_2,\,J_3,\,J_4$ for the orbital configuration of Table\,\ref{tavola3}.
Middle four rows: Uncancelled mismodeled precessions due to the zonal harmonics $J_5,\,J_6,\,J_7,\,J_8$, in $\mathrm{mas\,yr}^{-1}$, linearly combined according to \rfr{kombo}; the formal, statistical errors $\upsigma_{J_\ell}$ in  $J_\ell,\,\ell=5,\,6,\,7,\,8$ of the model Tongji-Grace02s \citep{2018JGRB..123.6111C}, quoted in Table\,\ref{tavola0}, were used.
Lower four rows: pN precessions, in $\mathrm{mas\,yr}^{-1}$, linearly combined according to \rfr{kombo}.
}\lb{tavola4bis}
\begin{center}
\small{
\begin{tabular}{|l|l|l|}
\hline
$c_1$ & $-3.91939$ & $-$\\
$c_2$ & $40.7154$ & $-$\\
$c_3$ & $2.20981$ & $-$\\
\hline
$\upsigma_{J_5}$ & $0.003$ & $\mathrm{mas\,yr}^{-1}$\\
$\upsigma_{J_6}$ & $0.002$ & $\mathrm{mas\,yr}^{-1}$\\
$\upsigma_{J_7}$ & $0.002$ & $\mathrm{mas\,yr}^{-1}$\\
$\upsigma_{J_8}$ & $0.001$ & $\mathrm{mas\,yr}^{-1}$\\
\hline
$J_2\,c^{-2}$ & $4.373$ & $\mathrm{mas\,yr}^{-1}$\\
$\varepsilon\,c^{-2}$ & $-0.001$ & $\mathrm{mas\,yr}^{-1}$\\
$S\,c^{-2}$ & $-9.952$ & $\mathrm{mas\,yr}^{-1}$\\
$M\,c^{-2}$ & $6,033.6$ & $\mathrm{mas\,yr}^{-1}$\\
\hline
\end{tabular}
}
\end{center}
\end{table}
\begin{table}
\caption{Numerically integrated nominal rates of change, averaged over one orbital revolution, of  the semimajor axis $a$, the eccentricity $e$, the inclination $I$, the longitude of the ascending node $\Omega$, the argument of pericenter $\omega$, and the mean anomaly at epoch $\eta$ induced by the neutral atmospheric drag for the orbital configuration of Table~\ref{tavola3}. The units are $\textrm{m}~\textrm{yr}^{-1}$ for $\ang{\dot a}$, and $\textrm{mas}~\textrm{yr}^{-1}$ for $\ang{\dot e},\,\ang{\dot I},\,\ang{\dot\Omega},\,\ang{\dot\omega},\,\ang{\dot\eta}$. For the satellite, assumed spherical in shape and passive, we adopted $C_\mathrm{D}=3.5,\,\Sigma=2.69\times 10^{-4}$ as for the existing LARES \citep{2017AcAau.140..469P}. In regard to the Earth's neutral atmospheric density, we adopted $r_0=r_\mathrm{min}=a\ton{1-e}=641.86\,\mathrm{km},\,\rho_0=\rho_\mathrm{max}=\ton{6.9-1.11}\times 10^{-14}\,\mathrm{kg\,m}^{-3},\,\rho_\mathrm{min}=0.0001\,\rho_\mathrm{L},\,\lambda=3,463.23\,\mathrm{km}-3,843.48\,\mathrm{km}$; the neutral atmospheric density at the  height of LAGEOS is $\rho_\mathrm{L}=6.579\times 10^{-18}\,\mathrm{kg\,m}^{-3}$ \citep{2015CQGra..32o5012L}. See Appendix\,\ref{appendrag} for details. Neither approximations in $e$ nor in $\nu\doteq \Psi/\nk$ were used. The value of $\rho_0=\rho_\mathrm{max}$ was kept fixed over one orbital revolution. Cfr. with the analytical plots in Figure\,\ref{fig3} and the numerically produced time series in Figure\,\ref{fig4}.
}\lb{tavola4ter}
\begin{center}
\small{
\begin{tabular}{|l|l|l|l|l|l|l|}
\hline
    $\rho_0\,\ton{\mathrm{kg\,m}^{-3}}$
  & $\ang{\dot a}\,\left(\textrm{m}\,\textrm{yr}^{-1}\right)$
  & $\ang{\dot e}\,\left(\textrm{mas}\,\textrm{yr}^{-1}\right)$
  & $\ang{\dot I}\,\left(\textrm{mas}\,\textrm{yr}^{-1}\right)$
  & $\ang{\dot\Omega}\,\left(\textrm{mas}\,\textrm{yr}^{-1}\right)$
  & $\ang{\dot\omega}\,\left(\textrm{mas}\,\textrm{yr}^{-1}\right)$
  & $\ang{\dot\eta}\,\left(\textrm{mas}\,\textrm{yr}^{-1}\right)$ \\
\hline
$6.9\times 10^{-14}$ & $-164.65$ & $-152.96$ & $-2.24$ & $0.69$ & $-0.30$ & $0.02$\\
$1.11\times 10^{-14}$ & $-27.6$ & $-25.6$ & $-0.41$ & $0.15$ & $-0.07$ & $0.008$\\
\hline
\end{tabular}
}
\end{center}
\end{table}
\begin{figure}
\begin{center}
\centerline{
\vbox{
\begin{tabular}{cc}
\epsfysize= 5.0 cm\epsfbox{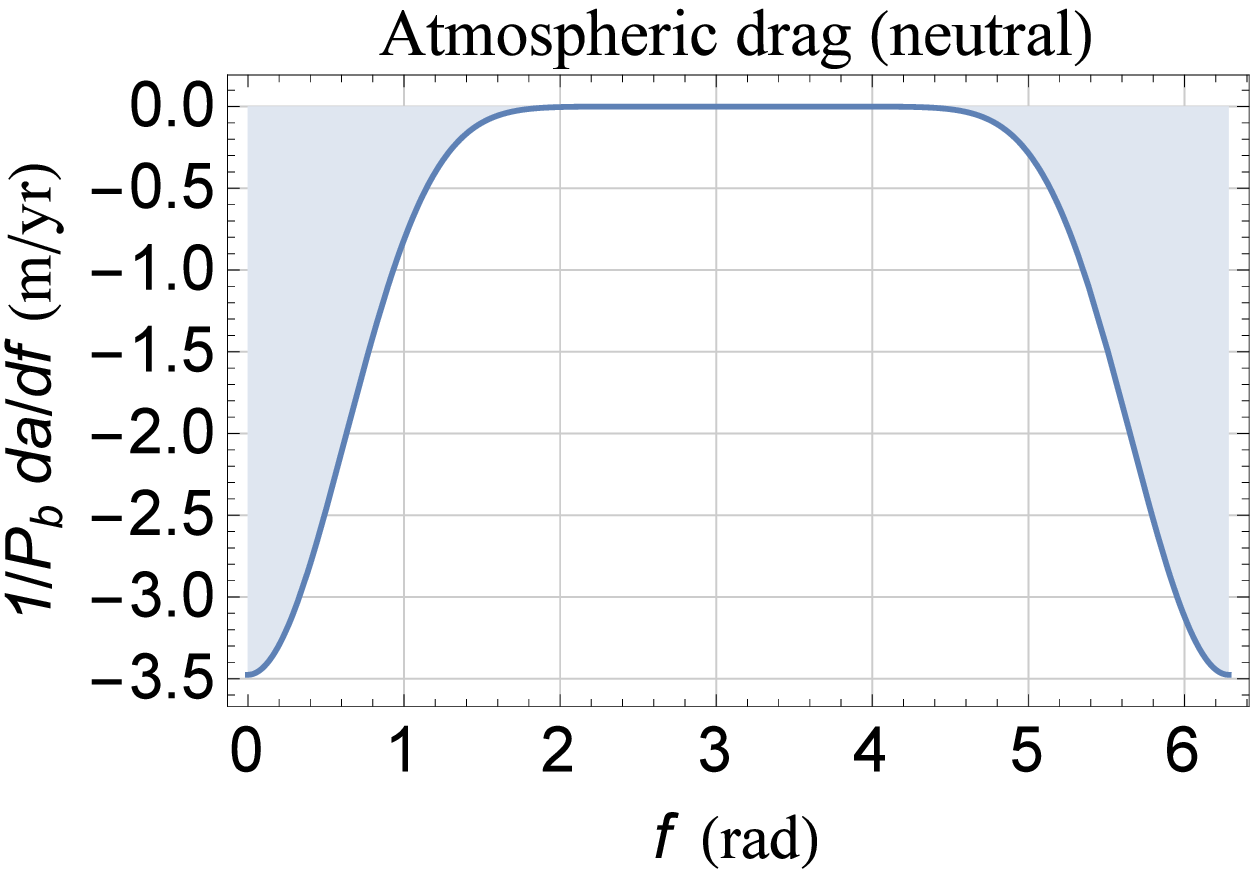}&
\epsfysize= 5.0 cm\epsfbox{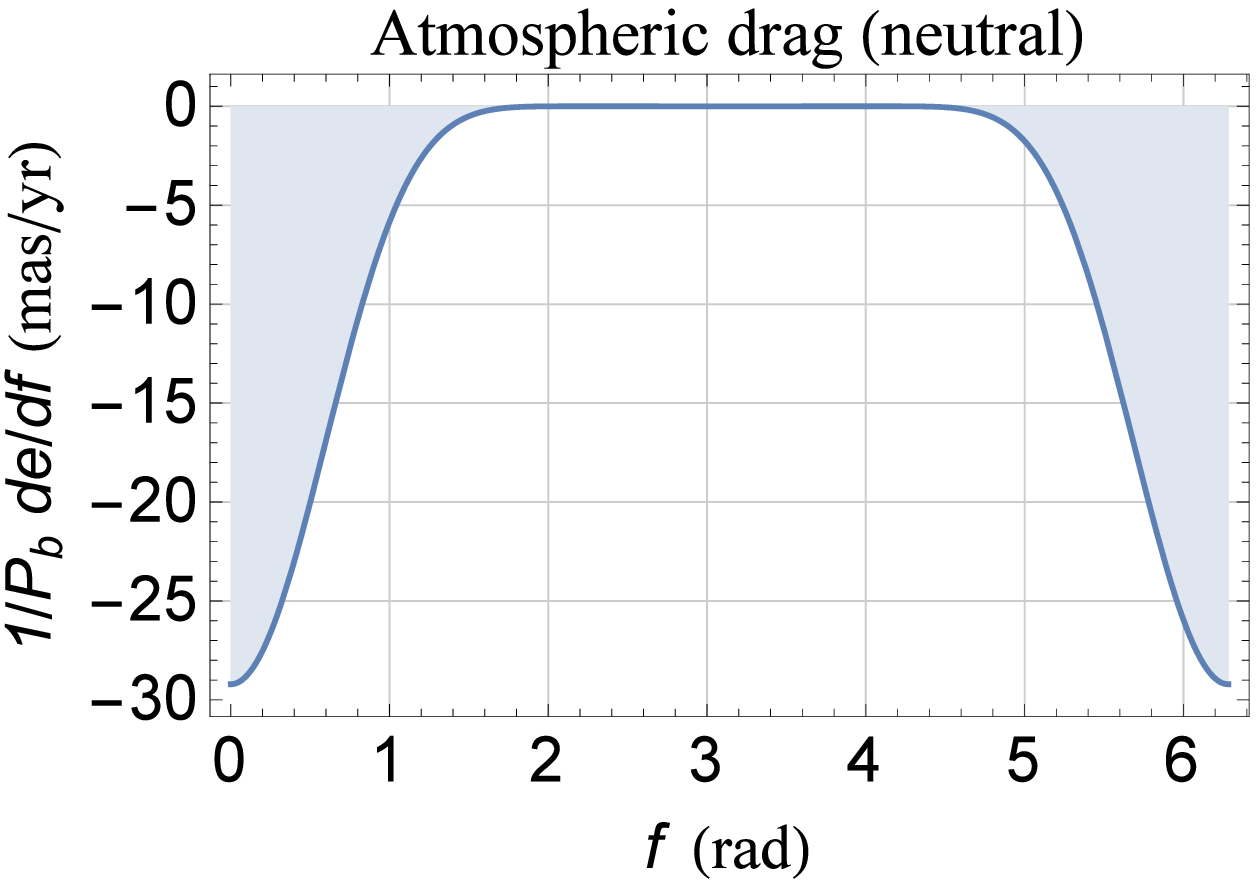}\\
\epsfysize= 5.0 cm\epsfbox{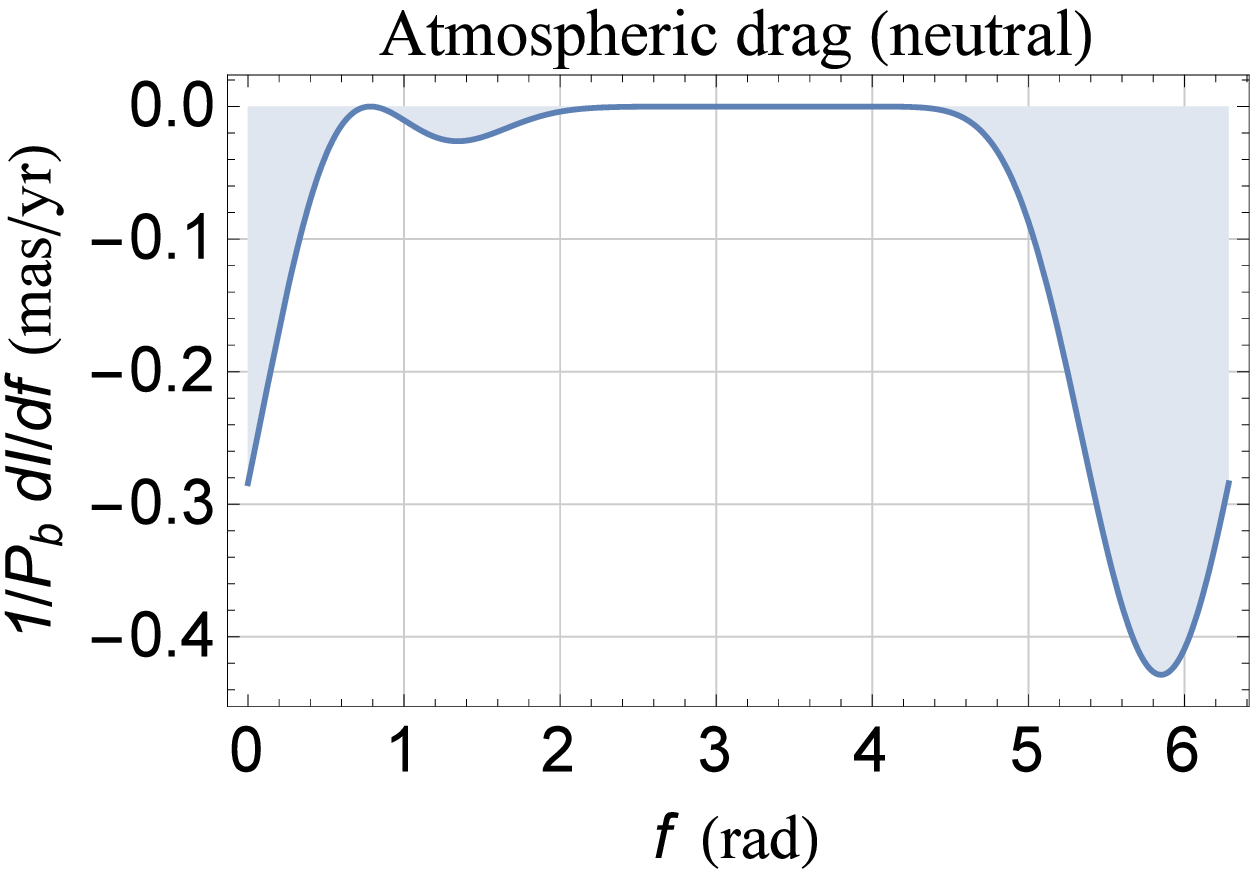}&
\epsfysize= 5.0 cm\epsfbox{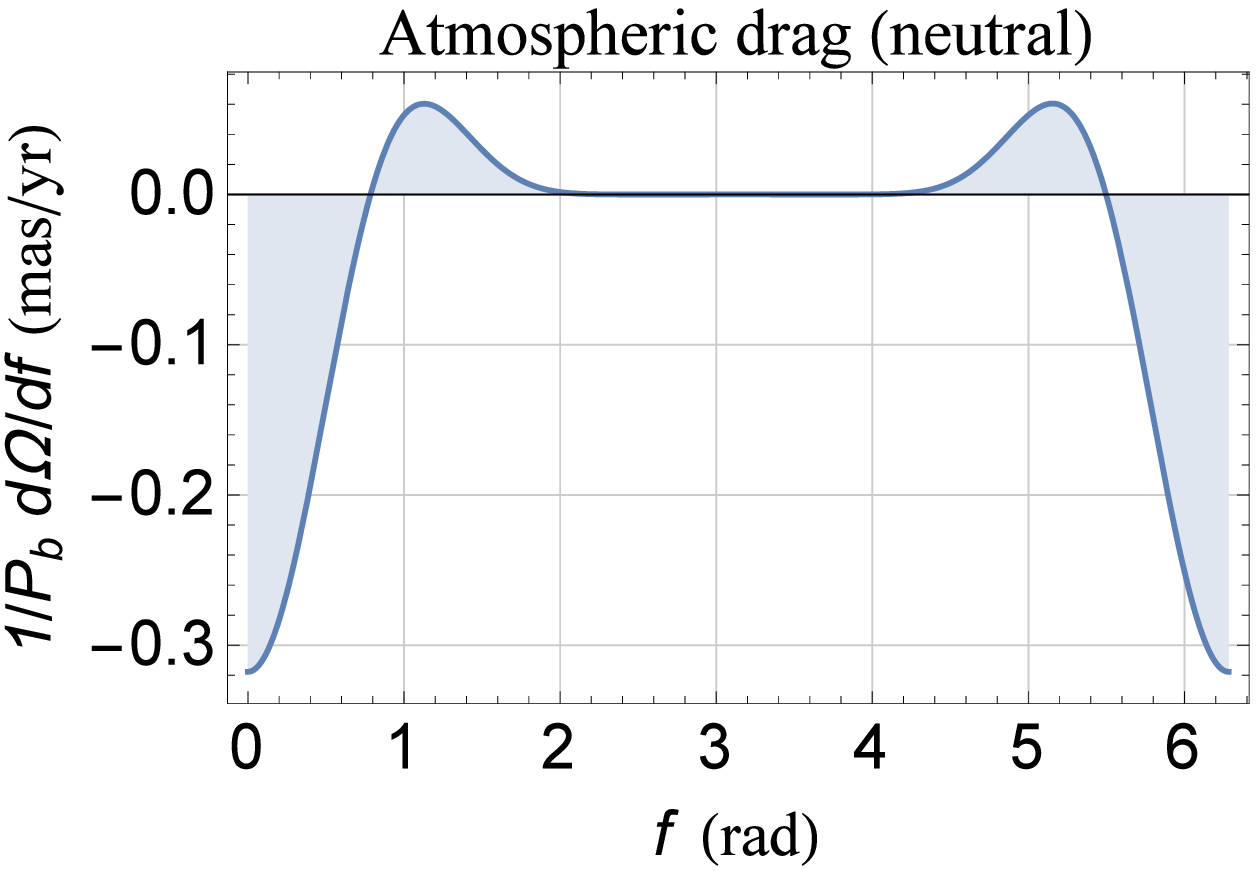}\\
\epsfysize= 5.0 cm\epsfbox{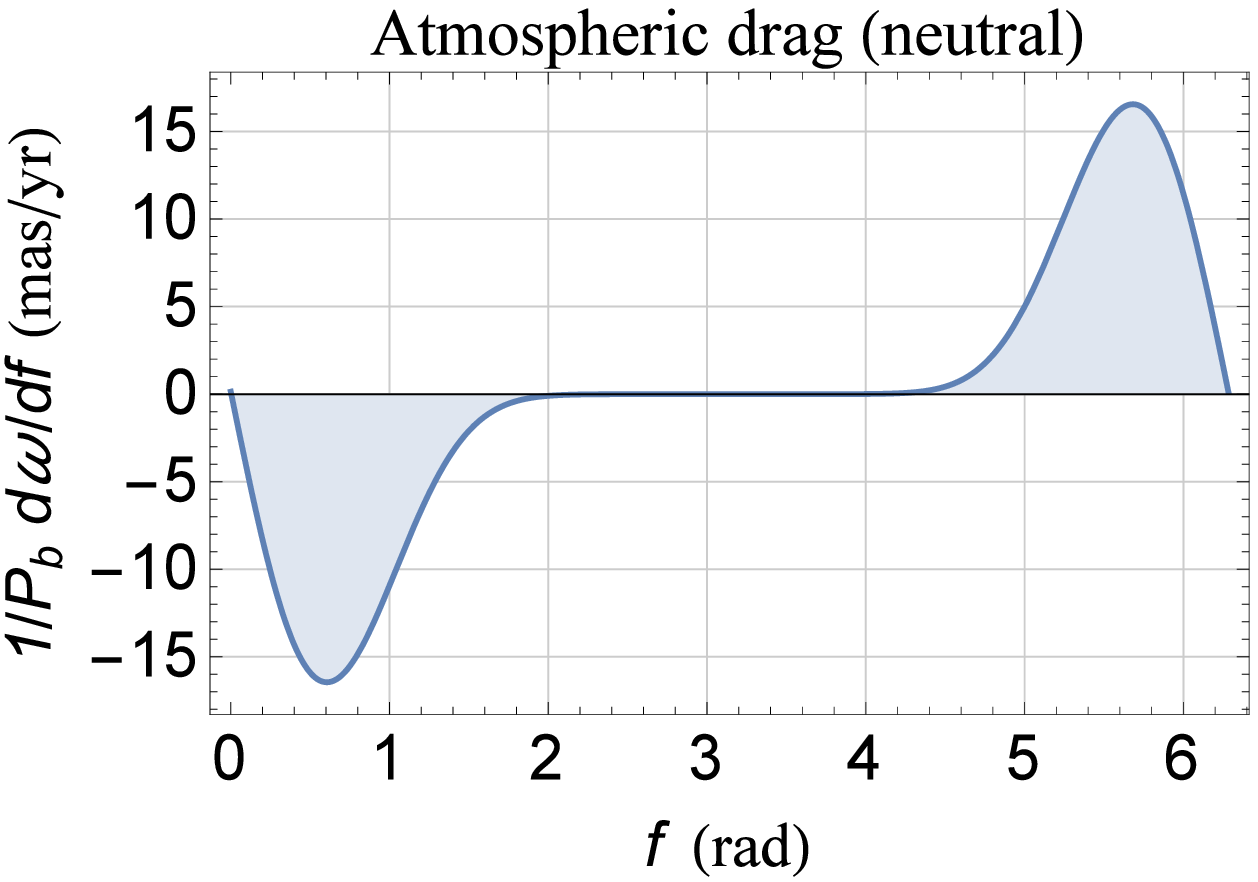}&
\epsfysize= 5.0 cm\epsfbox{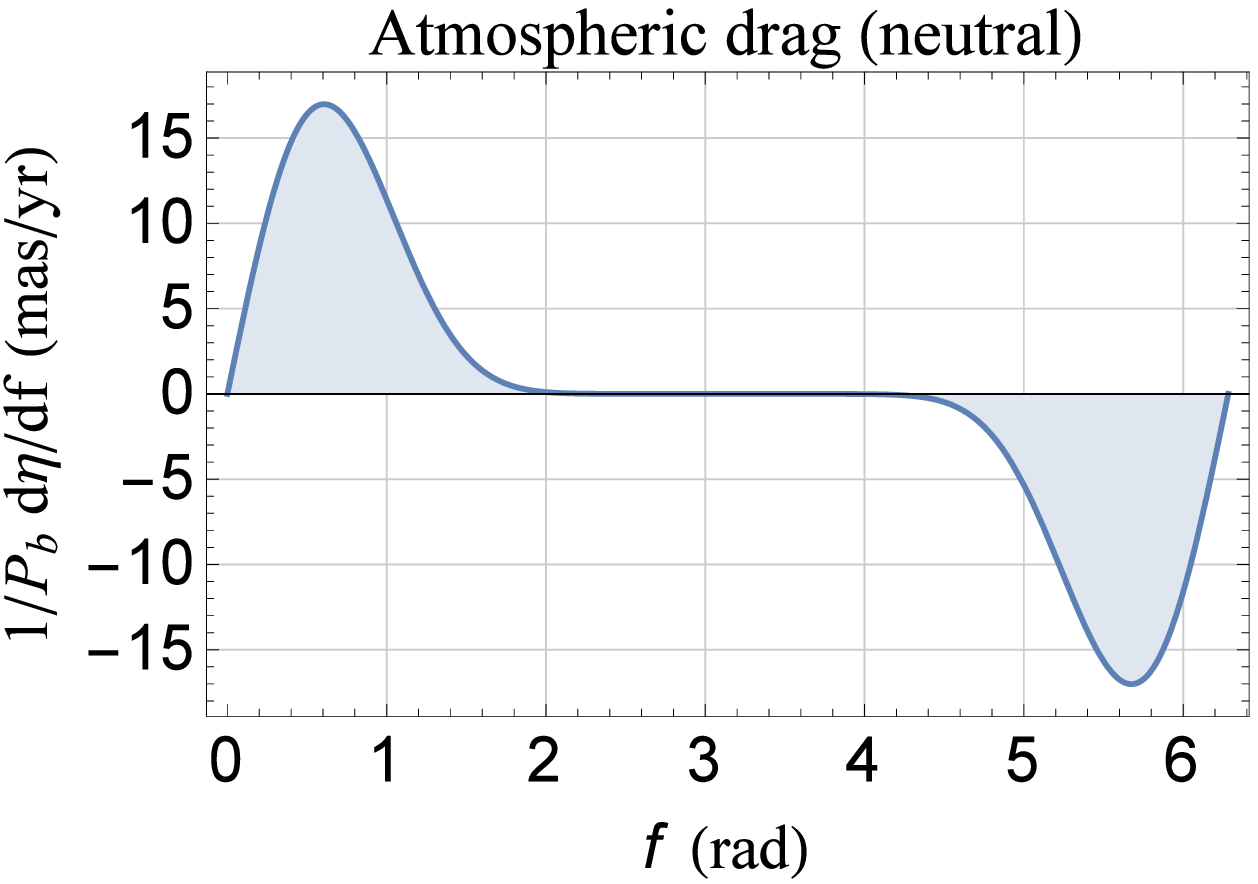}\\
\end{tabular}
}
}
\caption{Plots of \rfrs{uno}{sei} as functions of the true anomaly $f$ from $0$ to $2\uppi$ for the orbital configuration of Table\,\ref{tavola1}. For the satellite, assumed spherical in shape and passive, we adopted $C_\mathrm{D}=3.5,\,\Sigma=2.69\times 10^{-4}$ as for the existing LARES \citep{2017AcAau.140..469P}. In regard to the Earth's atmospheric density, we adopted $r_0=r_\mathrm{min}=a\ton{1-e}=1,046.86\,\mathrm{km},\,\rho_0=\rho_\mathrm{max}=7.3\times 10^{-15}\,\mathrm{kg\,m}^{-3},\,\lambda=872.87\,\mathrm{km}$. Neither approximations in $e$ nor in $\nu\doteq \Psi/\nk$ were used.  The areas of the regions delimited by the curves and the $f$ axis are the rates of change of the orbital elements averaged over one orbital period $\Pb$; they are numerically calculated and displayed in Table\,\ref{tavola2ter}. The value of $\rho_0=\rho_\mathrm{max}$ was kept fixed over one orbital revolution. }\label{fig1}
\end{center}
\end{figure}
\begin{figure*}
\begin{center}
\centerline{
\vbox{
\begin{tabular}{cc}
\epsfysize= 5.0 cm\epsfbox{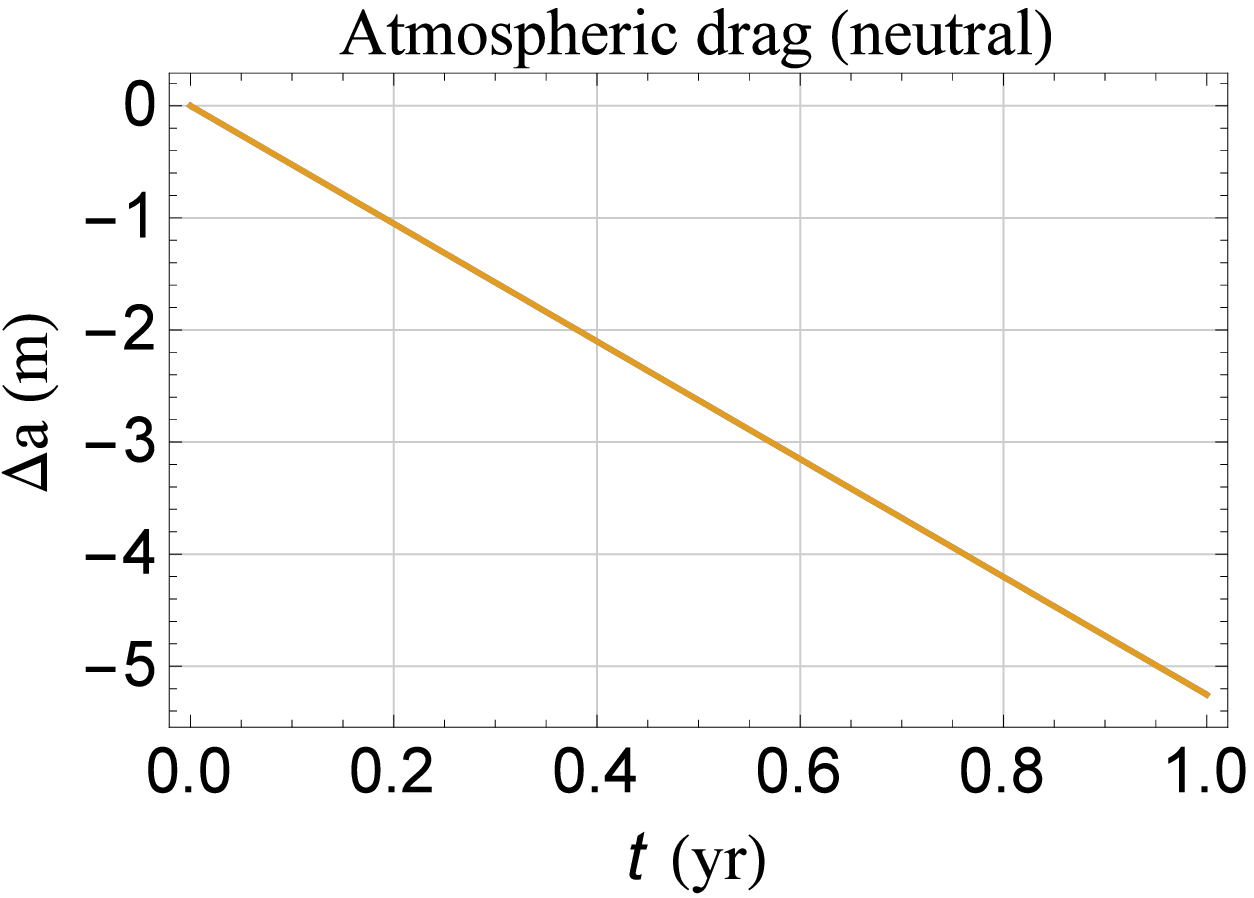}&
\epsfysize= 5.0 cm\epsfbox{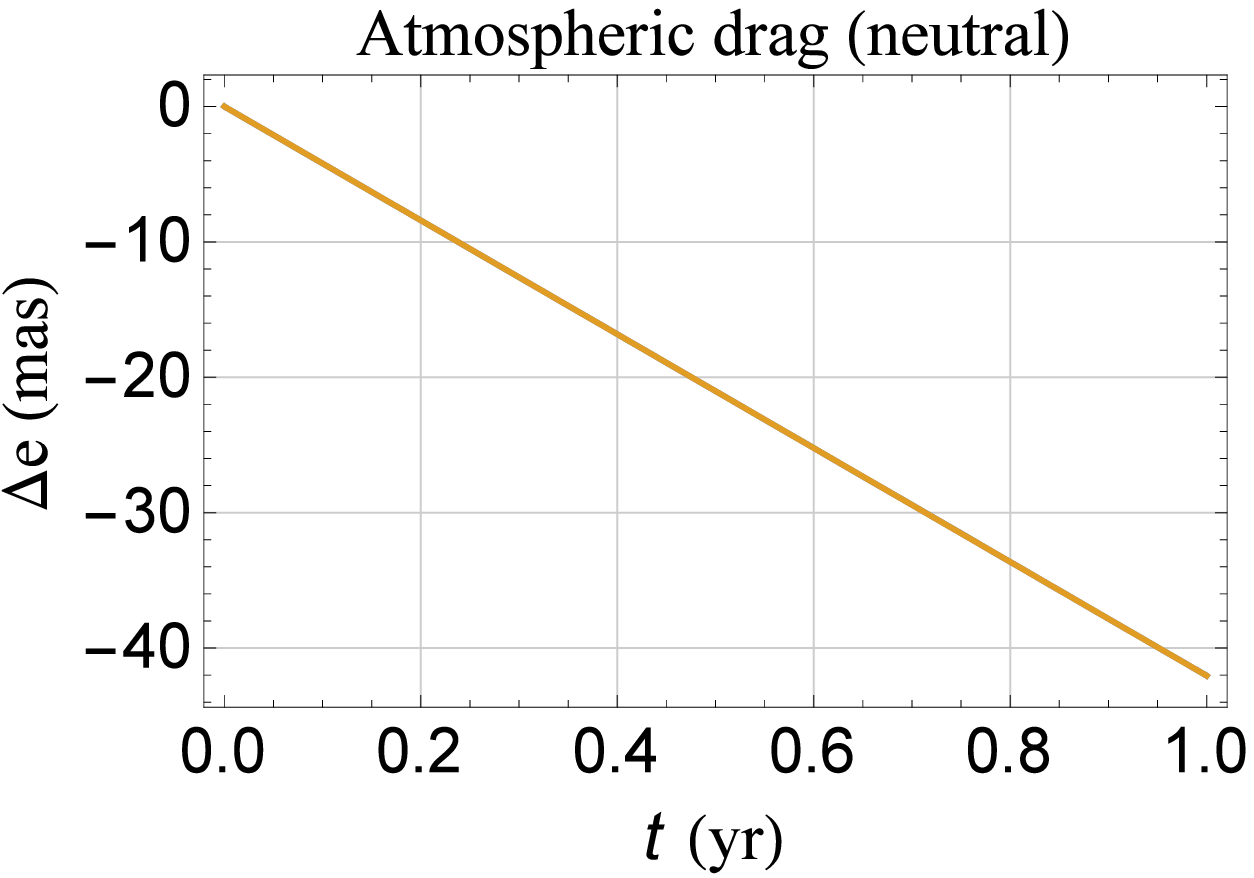}\\
\epsfysize= 5.0 cm\epsfbox{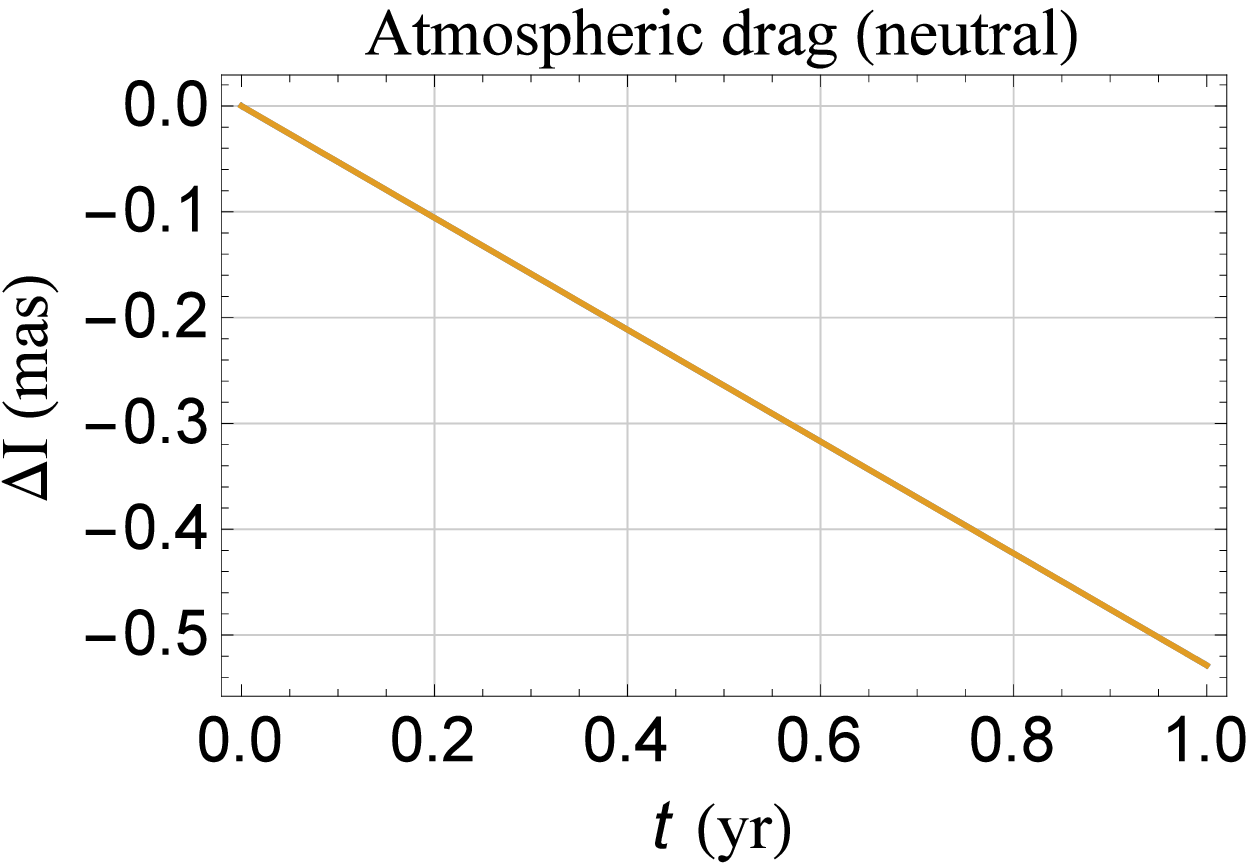}&
\epsfysize= 5.0 cm\epsfbox{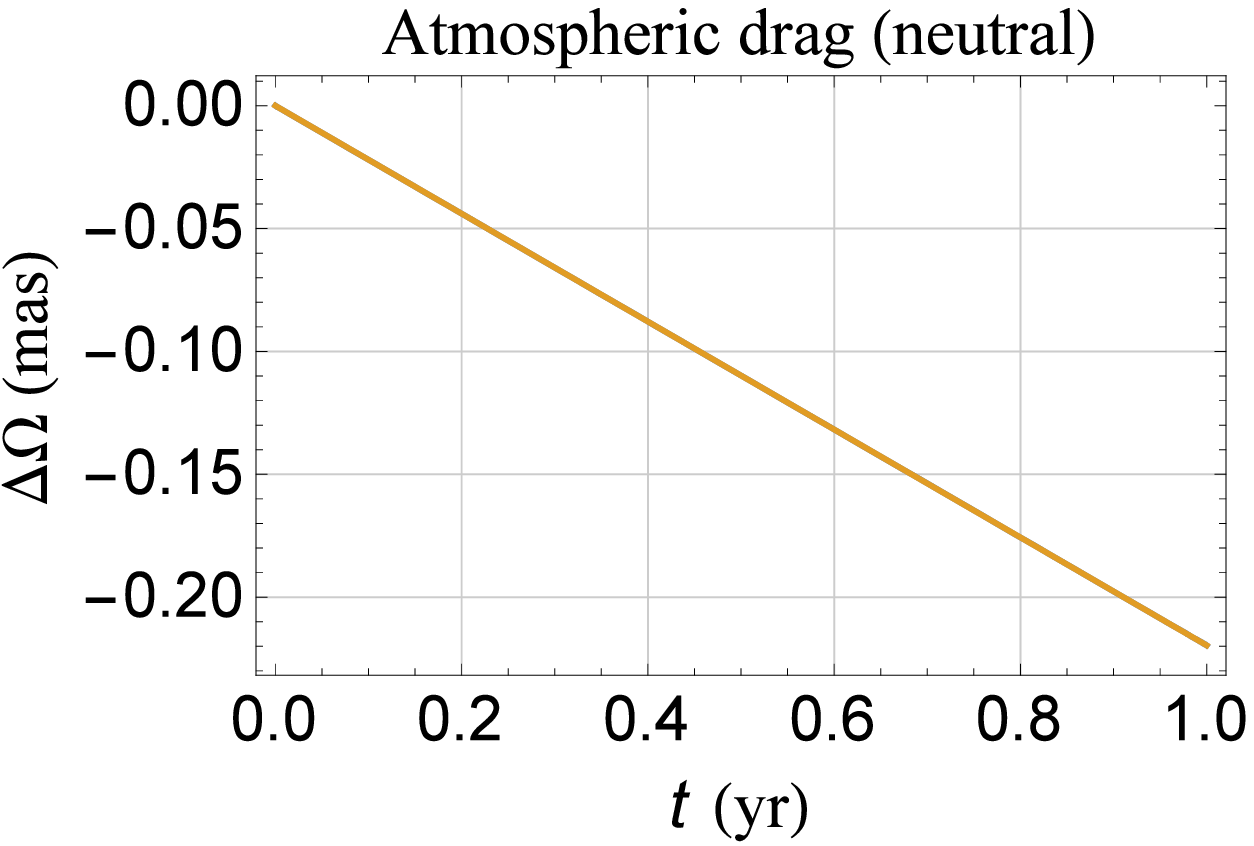}\\
\epsfysize= 5.0 cm\epsfbox{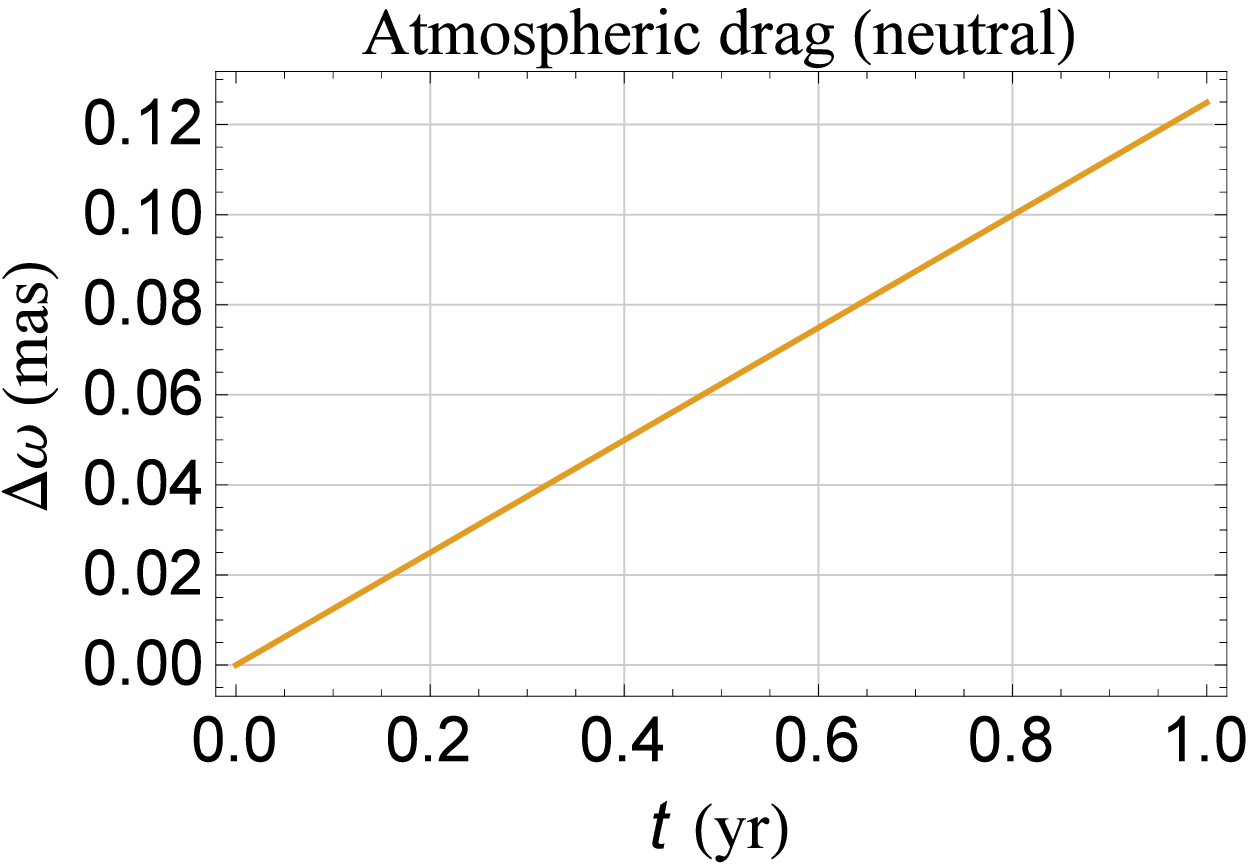}&
\epsfysize= 5.0 cm\epsfbox{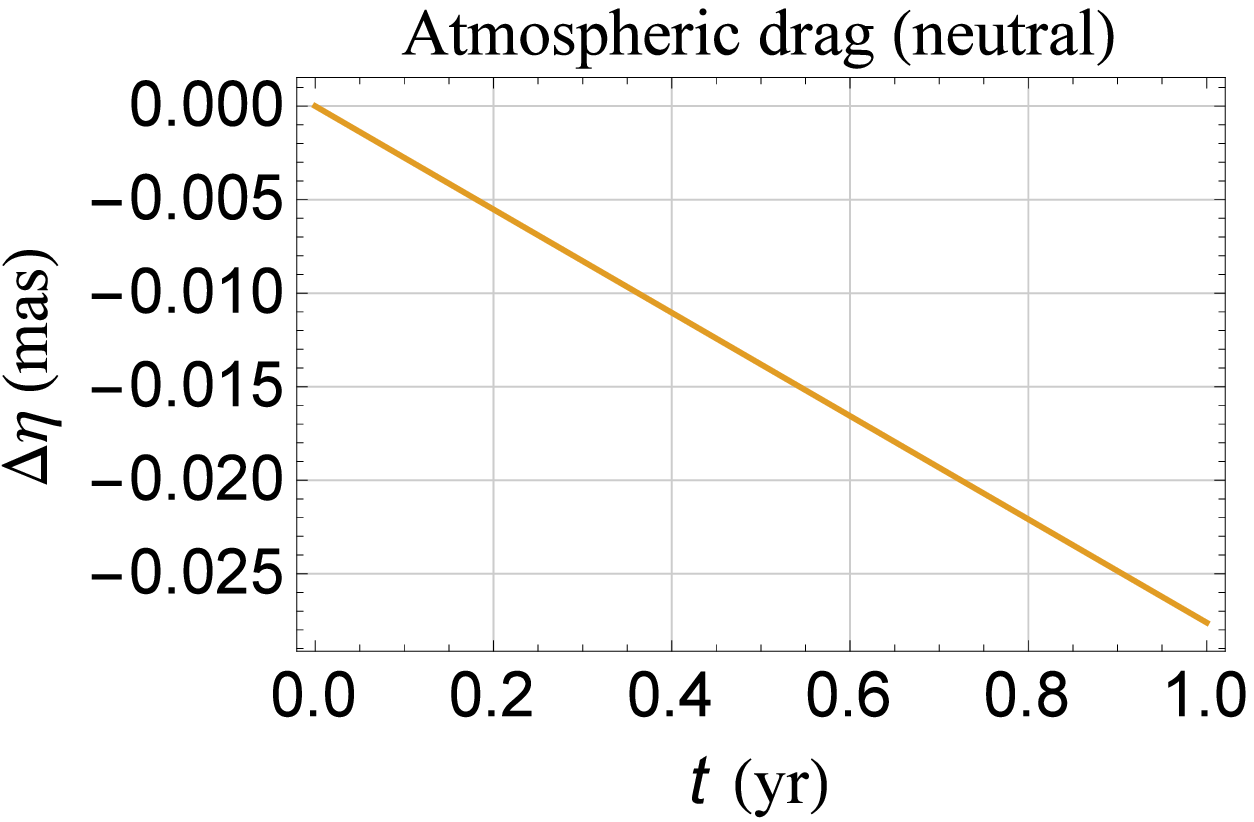}\\
\end{tabular}
}
}
\caption{Numerically integrated shifts of the semimajor axis $a$, the eccentricity $e$, the longitude of the ascending node $\Omega$, the argument of the perigee $\omega$, and the mean anomaly at epoch $\eta$   for the orbital configuration of Table\,\ref{tavola1}. For the satellite, assumed spherical in shape and passive, we adopted $C_\mathrm{D}=3.5,\,\Sigma=2.69\times 10^{-4}$ as for the existing LARES \citep{2017AcAau.140..469P}. In regard to the Earth's atmospheric density, we adopted $r_0=r_\mathrm{min}=a\ton{1-e}=1,046.86\,\mathrm{km},\,\rho_0=\rho_\mathrm{max}=7.3\times 10^{-15}\,\mathrm{kg\,m}^{-3},\,\lambda=872.87\,\mathrm{km}$. Cfr. with the semi-analytical results of Figure\,\ref{fig1} and Table\,\ref{tavola2ter}.}\label{fig2}
\end{center}
\end{figure*}
\begin{figure}
\begin{center}
\centerline{
\vbox{
\begin{tabular}{cc}
\epsfysize= 5.0 cm\epsfbox{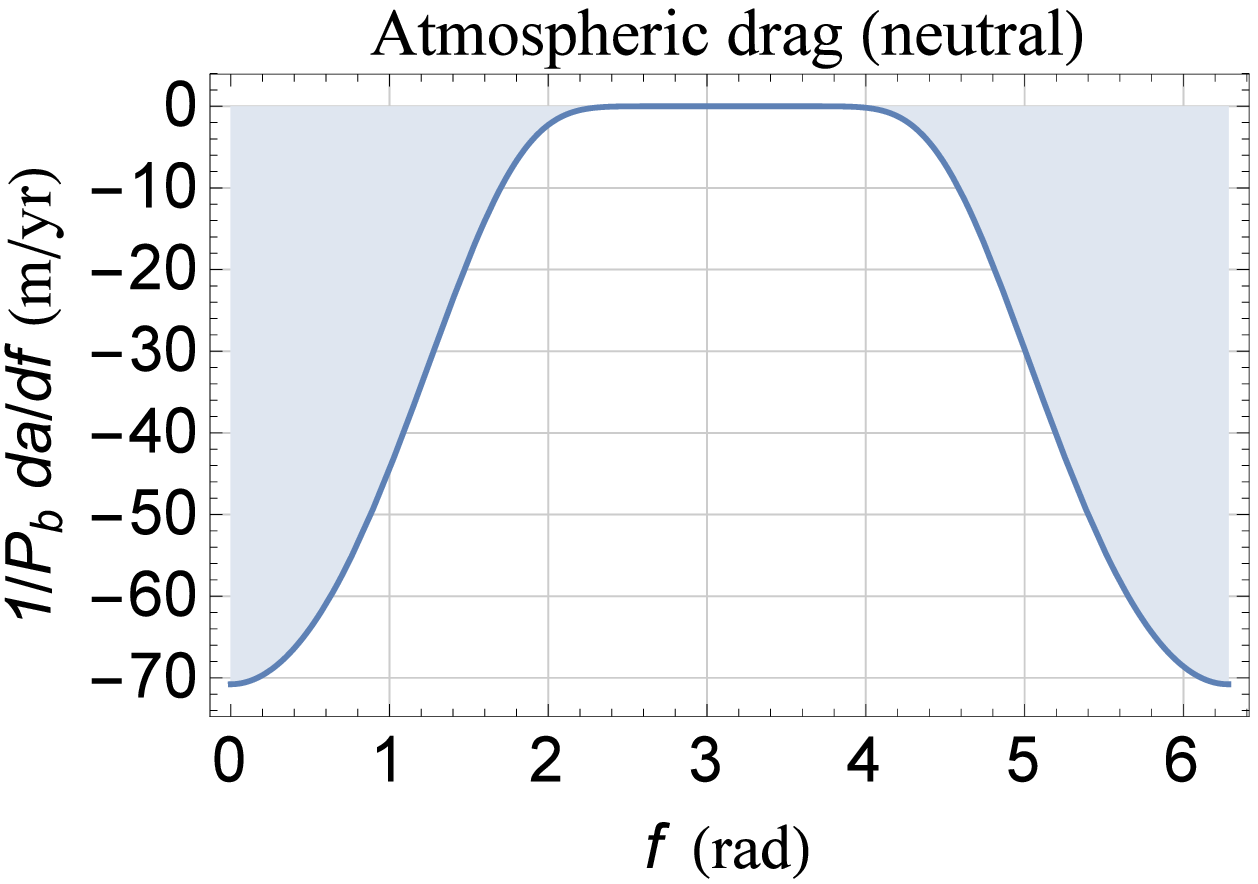}&
\epsfysize= 5.0 cm\epsfbox{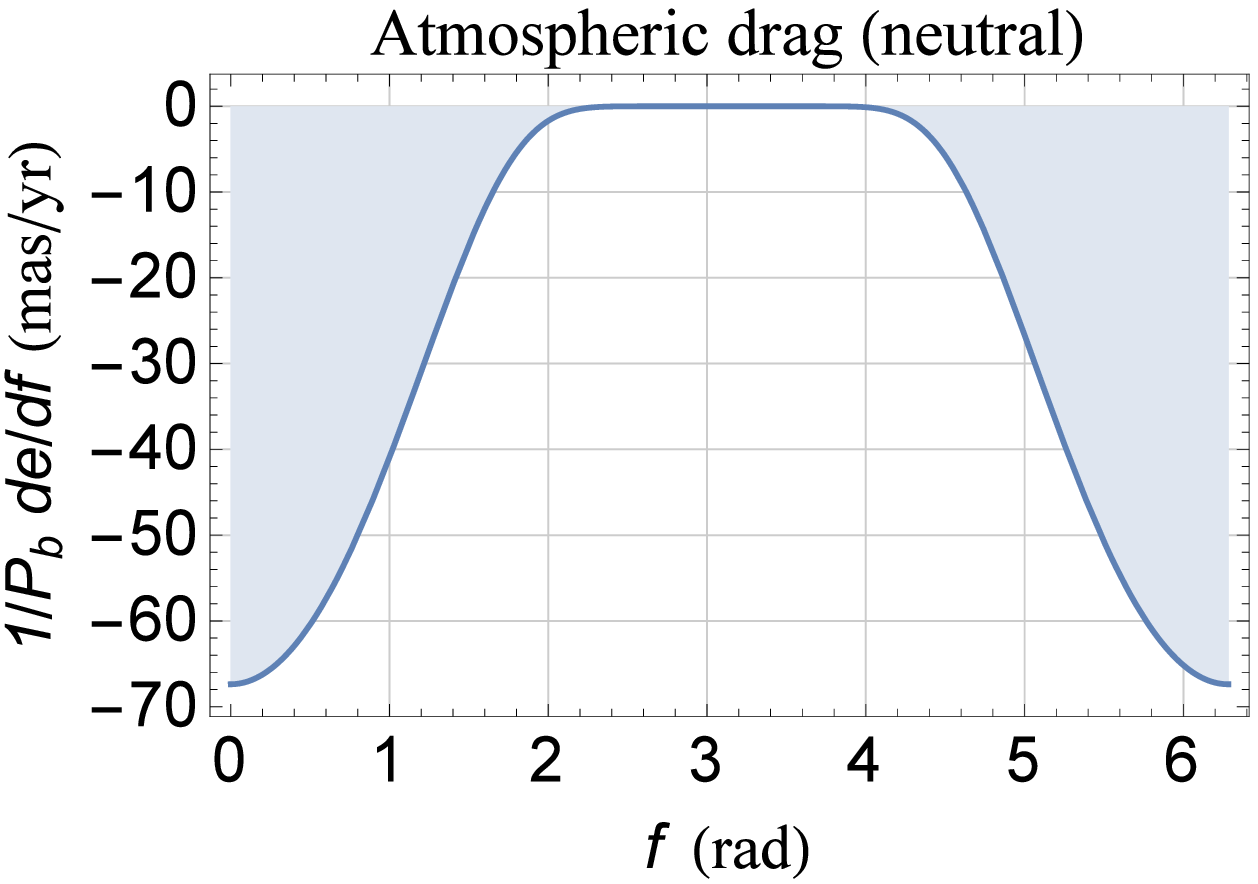}\\
\epsfysize= 5.0 cm\epsfbox{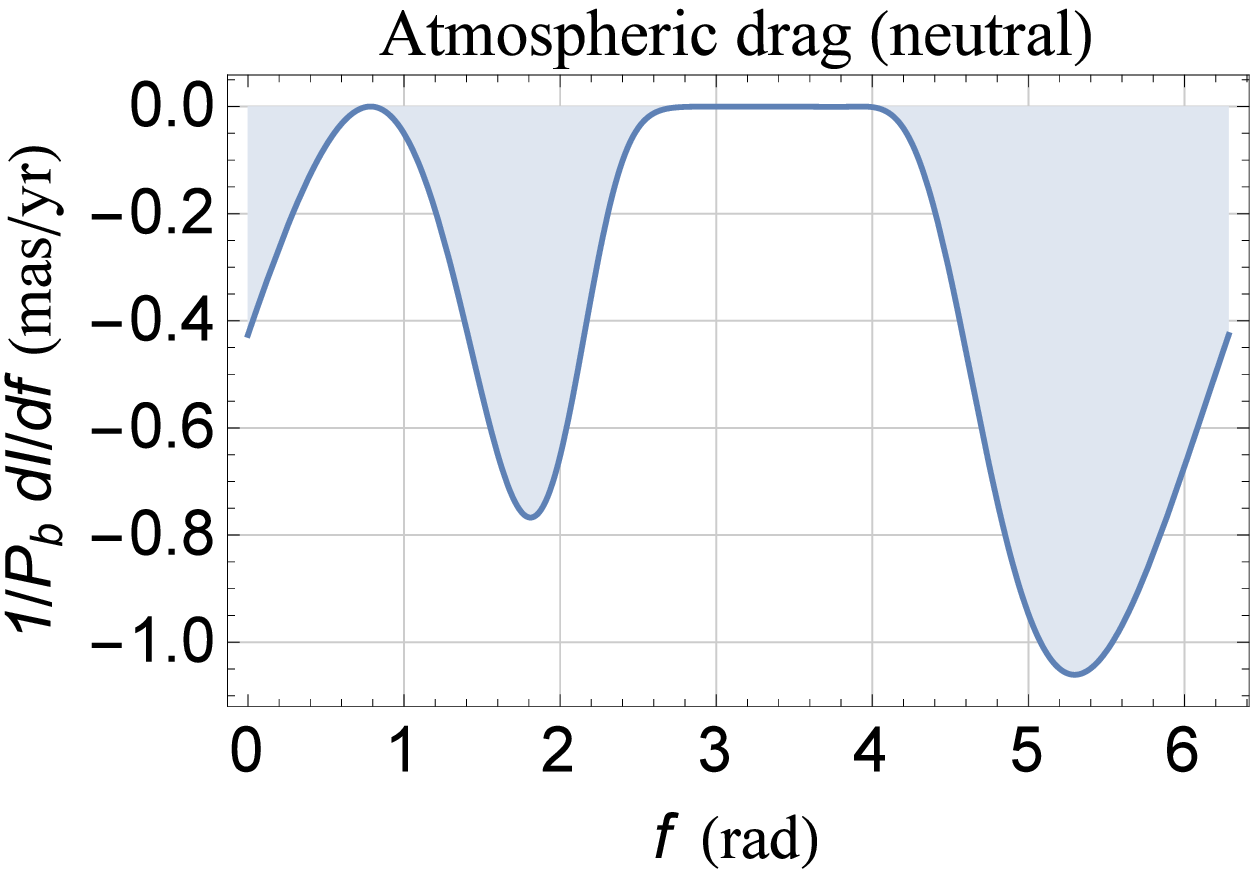}&
\epsfysize= 5.0 cm\epsfbox{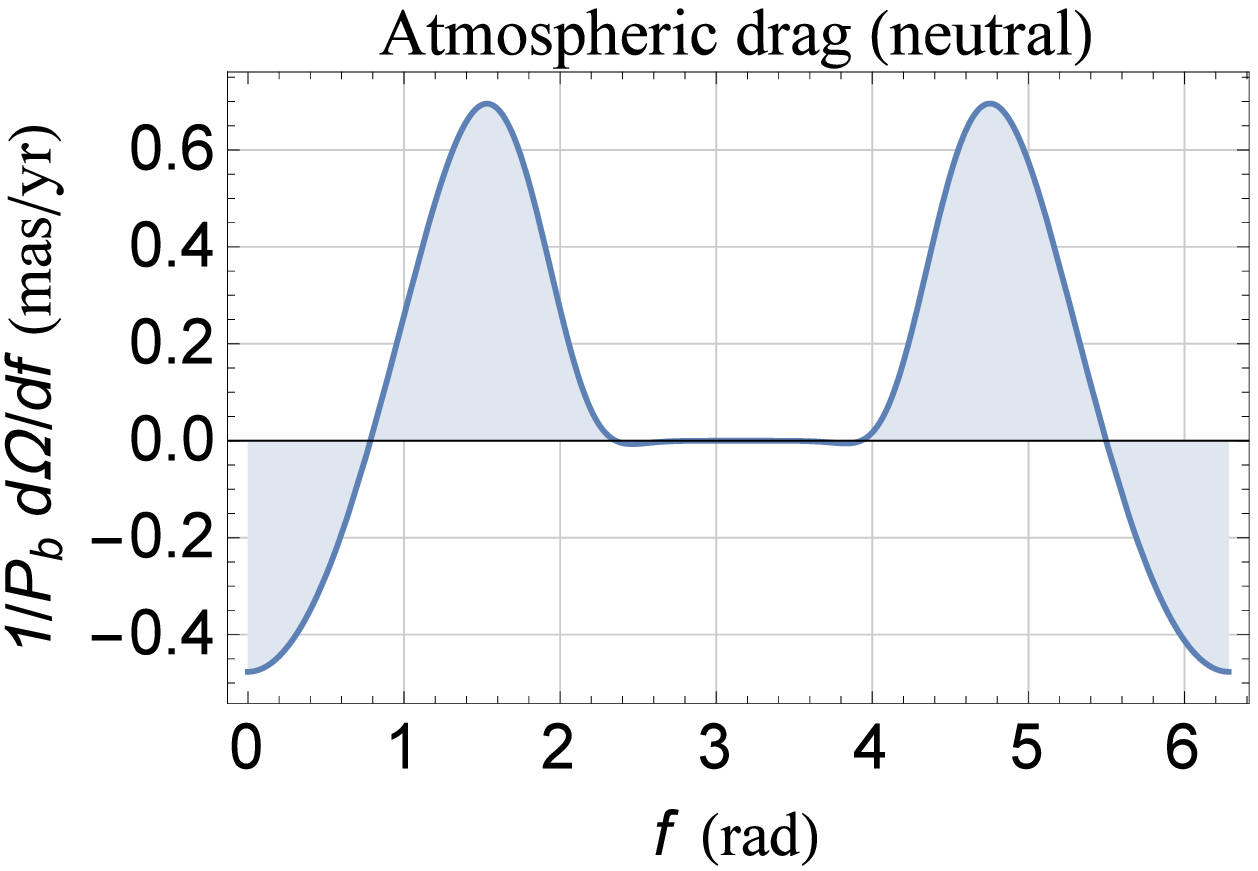}\\
\epsfysize= 5.0 cm\epsfbox{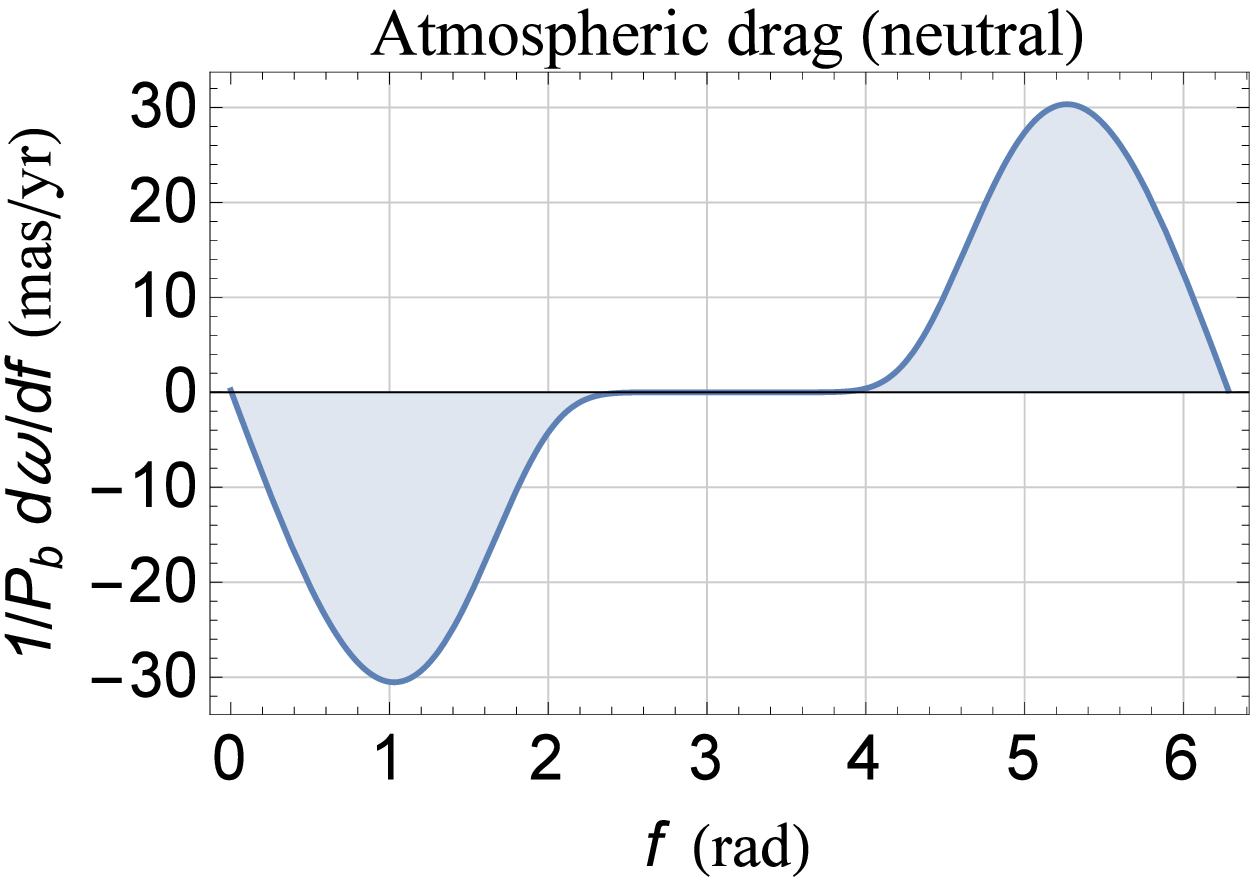}&
\epsfysize= 5.0 cm\epsfbox{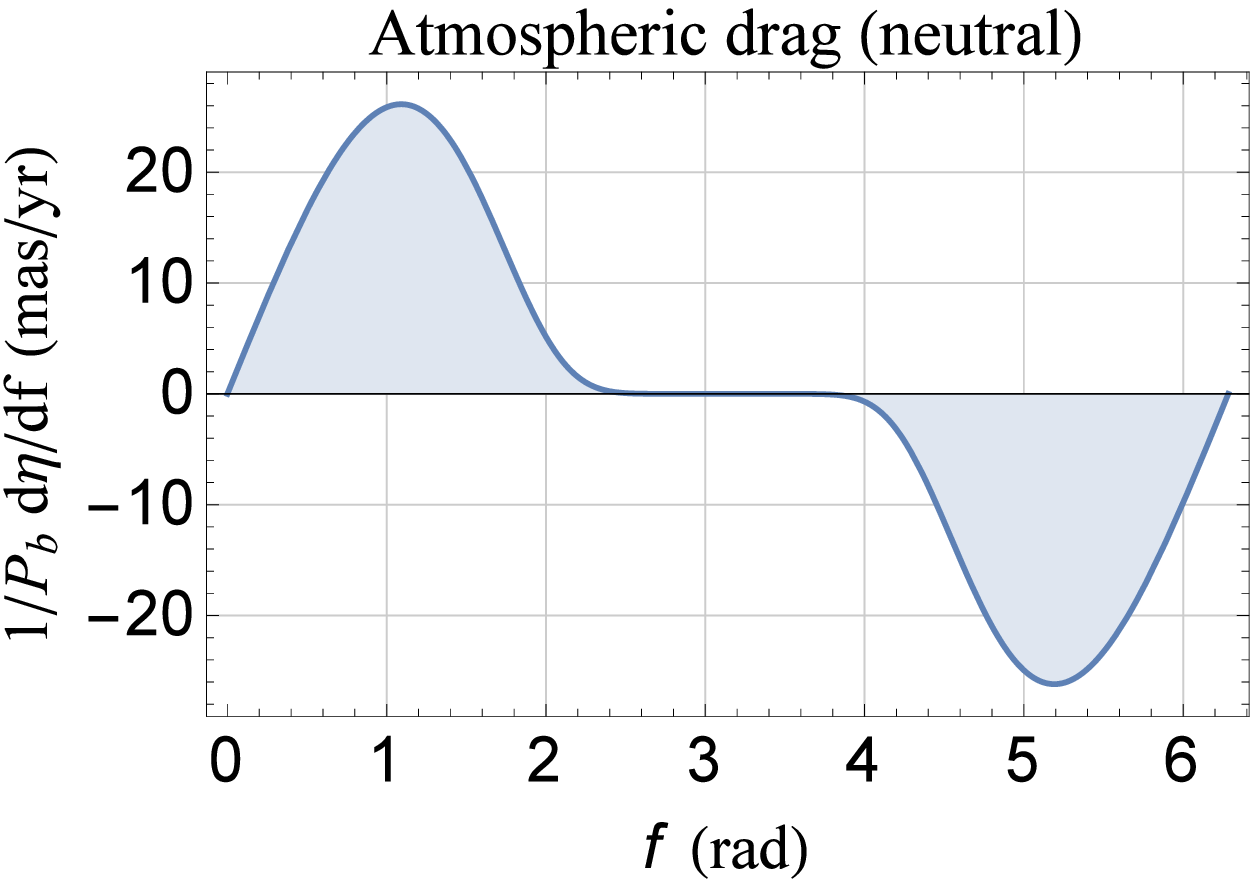}\\
\end{tabular}
}
}
\caption{Plots of \rfrs{uno}{sei} as functions of the true anomaly $f$ from $0$ to $2\uppi$ for the orbital configuration of Table\,\ref{tavola3}. For the satellite, assumed spherical in shape and passive, we adopted $C_\mathrm{D}=3.5,\,\Sigma=2.69\times 10^{-4}$ as for the existing LARES \citep{2017AcAau.140..469P}. In regard to the Earth's atmospheric density, we adopted $r_0=r_\mathrm{min}=a\ton{1-e}=641.86\,\mathrm{km},\,\rho_0=\rho_\mathrm{max}=6.9\times 10^{-14}\,\mathrm{kg\,m}^{-3},\,\lambda=3,463.23\,\mathrm{km}$. Neither approximations in $e$ nor in $\nu\doteq \Psi/\nk$ were used.  The areas of the regions delimited by the curves and the $f$ axis are the rates of change of the orbital elements averaged over one orbital period $\Pb$; they are numerically calculated and displayed in Table\,\ref{tavola4ter}. The value of $\rho_0=\rho_\mathrm{max}$ was kept fixed over one orbital revolution. }\label{fig3}
\end{center}
\end{figure}
\begin{figure*}
\begin{center}
\centerline{
\vbox{
\begin{tabular}{cc}
\epsfysize= 5.0 cm\epsfbox{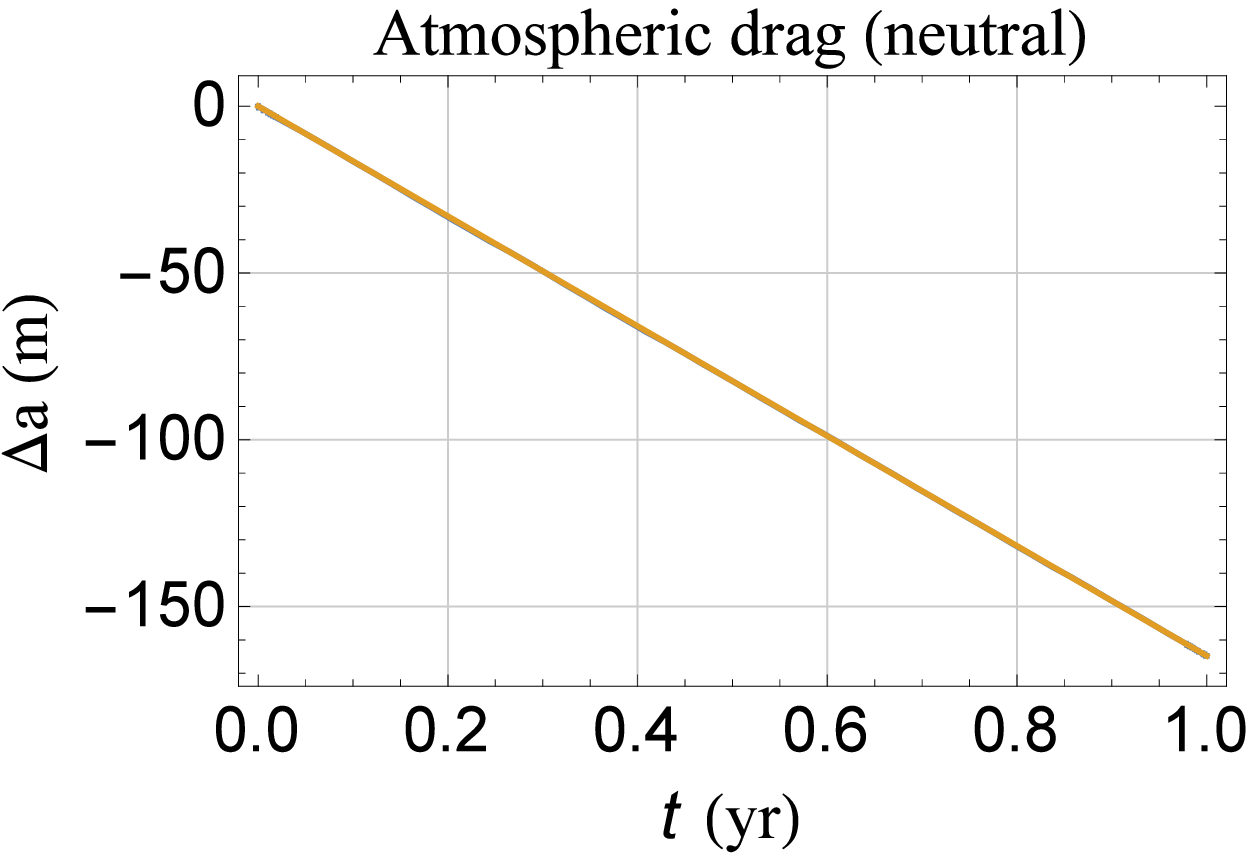}&
\epsfysize= 5.0 cm\epsfbox{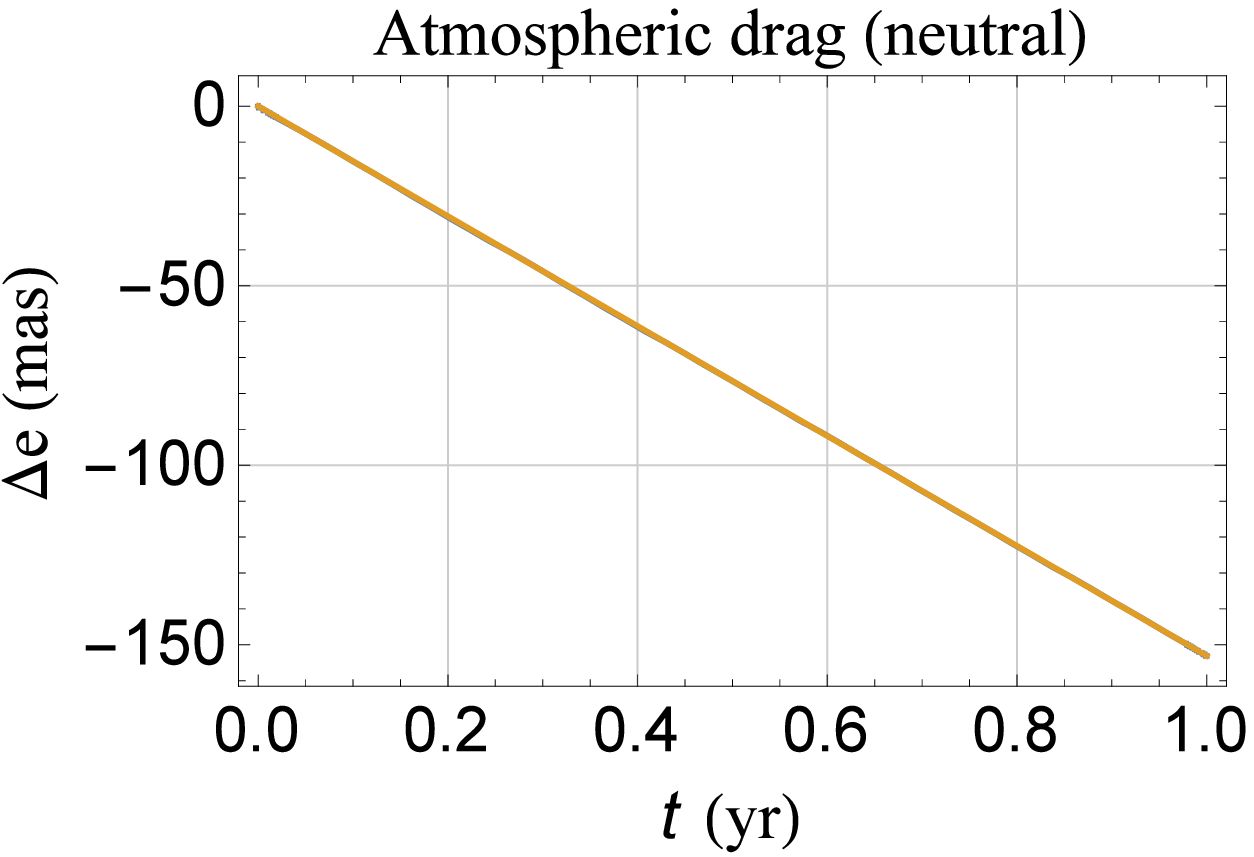}\\
\epsfysize= 5.0 cm\epsfbox{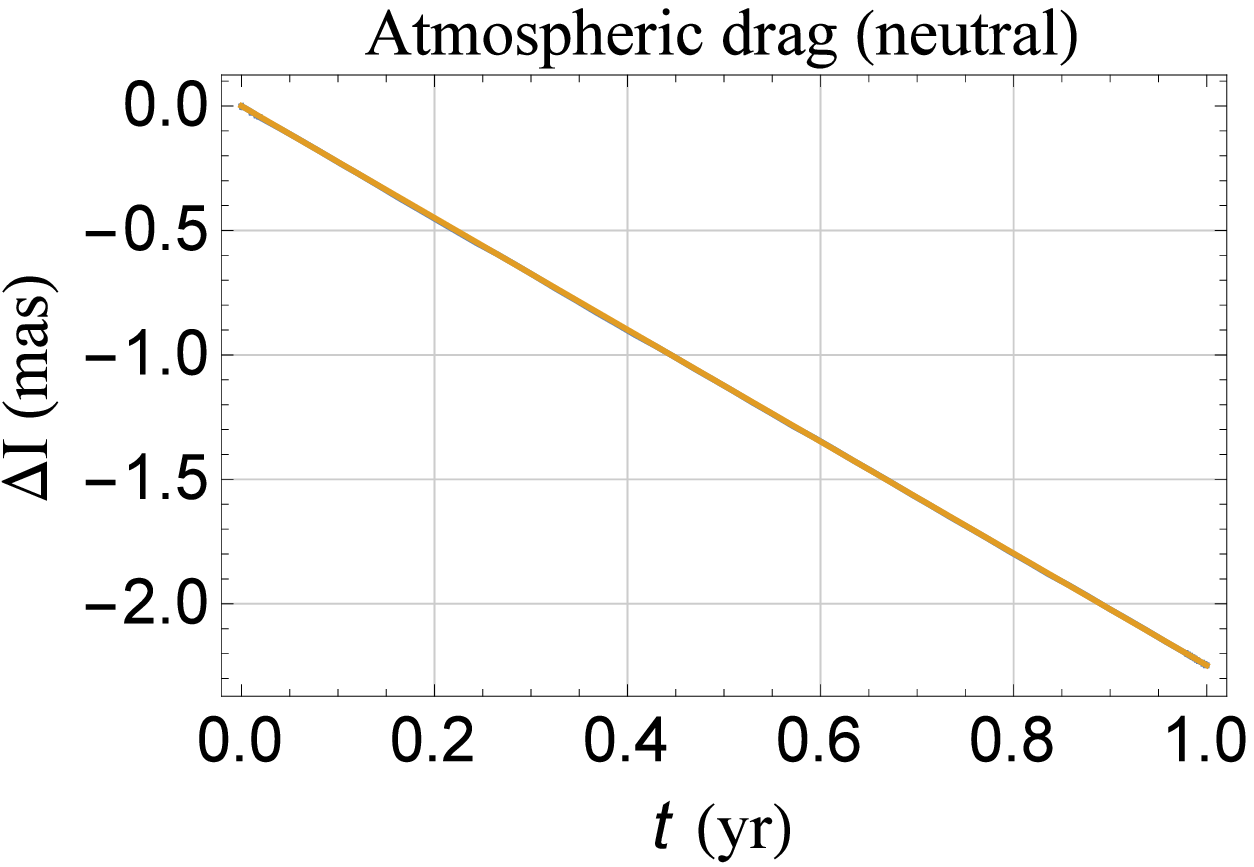}&
\epsfysize= 5.0 cm\epsfbox{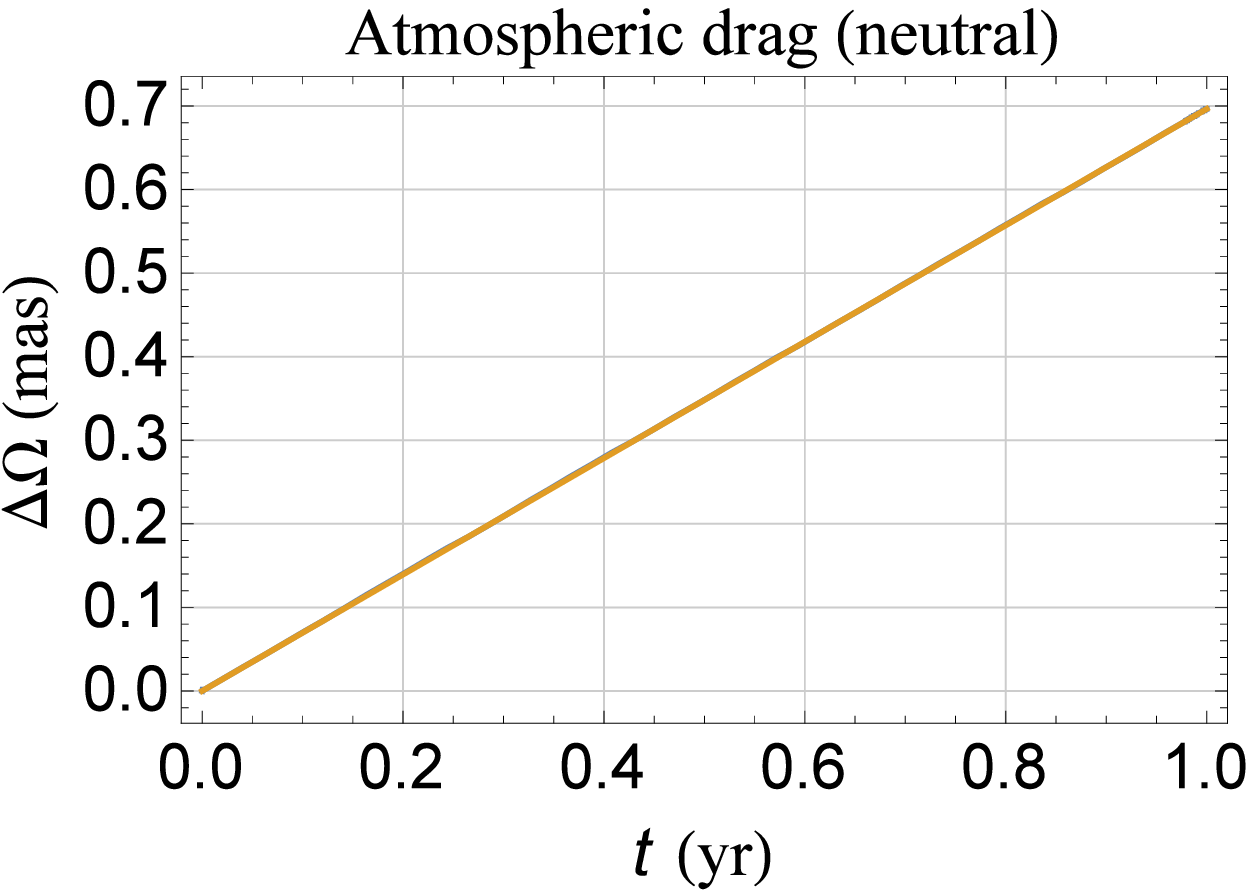}\\
\epsfysize= 5.0 cm\epsfbox{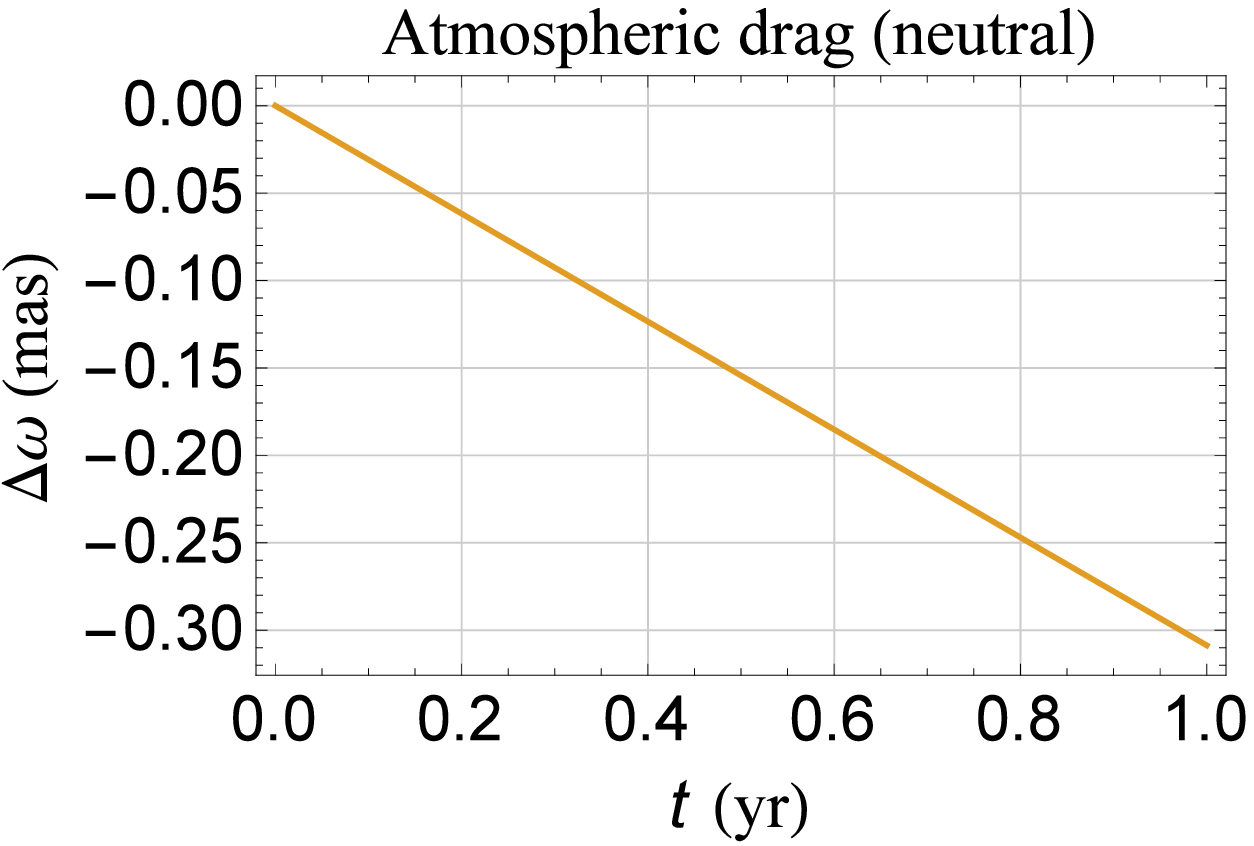}&
\epsfysize= 5.0 cm\epsfbox{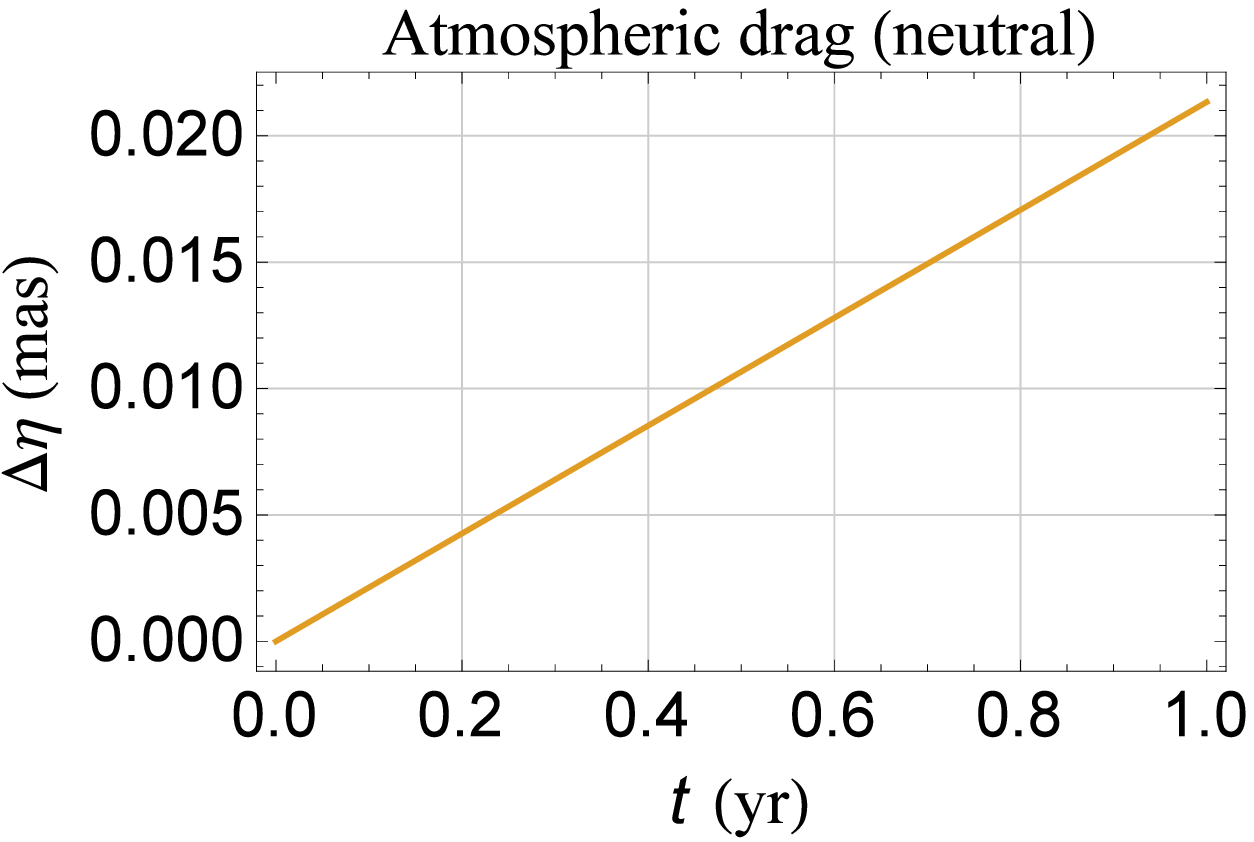}\\
\end{tabular}
}
}
\caption{Numerically integrated shifts of the semimajor axis $a$, the eccentricity $e$, the longitude of the ascending node $\Omega$, the argument of the perigee $\omega$, and the mean anomaly at epoch $\eta$   for the orbital configuration of Table\,\ref{tavola3}. For the satellite, assumed spherical in shape and passive, we adopted $C_\mathrm{D}=3.5,\,\Sigma=2.69\times 10^{-4}$ as for the existing LARES \citep{2017AcAau.140..469P}. In regard to the Earth's atmospheric density, we adopted $r_0=r_\mathrm{min}=a\ton{1-e}=641.86\,\mathrm{km},\,\rho_0=\rho_\mathrm{max}=6.9\times 10^{-14}\,\mathrm{kg\,m}^{-3},\,\lambda=3,463.23\,\mathrm{km}$. Cfr. with the semi-analytical results of Figure\,\ref{fig3} and Table\,\ref{tavola4ter}.}\label{fig4}
\end{center}
\end{figure*}
\clearpage
\section{Classical long-term rates of change of the Keplerian orbital elements up to degree $\ell=8$}\lb{appenb}
\renewcommand{\theequation}{B\arabic{equation}}
\setcounter{equation}{0}
Here, we will analytically work out the coefficients $\dot\kappa_{.\ell}\doteq\partial\ang{\dot\kappa}_{J_\ell}/\partial J_\ell,\,\kappa=e,\,I,\,\Omega,\,\omega,\,\eta$ of the long-term rates of change of the eccentricity $e$, the inclination $I$, the longitude of the ascending node $\Omega$, the argument of perigee $\omega$, and the mean anomaly at epoch $\eta$ induced by the first seven zonal harmonics of the geopotential up to degree $\ell=8$; the averaged rates of change of the semimajor axis $a$ are all vanishing. We will adopt the Lagrange planetary equations \citep{Bertotti03,2005som..book.....C,2008orbi.book.....X,2011rcms.book.....K} applied to the perturbing potential of degree $\ell$
\eqi
\Delta U_\ell\ton{\bds r} = \rp{\mu}{r}\,\ton{\rp{R_\mathrm{e}}{r}}^\ell\,J_\ell\,\mathcal{P}_\ell\ton{\xi},\,\ell=2,\,3,\ldots 8
\eqf
where $\mathcal{P}_\ell\ton{\xi}$ is the Legendre polynomial of degree $\ell$, averaged over one orbital period as disturbing function.
We will not make any a-priory assumption on the orbital configuration of the satellite. As far as the orientation of the Earth's spin axis, we will align it to the reference $z$ axis of an equatorial coordinate system. The following formulas include both the genuine secular and the harmonic components with the frequency of the perigee and its integer multiples.
\subsection{The eccentricity}
\begin{align}
\dot e_{.2} &= 0, \\ \nonumber \\
\dot e_{.3} &= -\rp{3\,\nk\,R_\mathrm{e}^3\,\ton{3 + 5\,\cos 2  I }\,\sin I\,\cos \omega}{16\,a^3\,\ton{1 -\,e^2}^2}, \\ \nonumber \\
\dot e_{.4} &= -\rp{15\,e\,\nk\,R_\mathrm{e}^4\,\ton{5 + 7\,\cos 2  I }\,\sin^2 I\,\sin 2 \omega  }{64\,a^4\,\ton{1 -\,e^2}^3}, \\ \nonumber \\
\dot e_{.5} \nonumber &= \rp{15\,\nk\,R_\mathrm{e}^5}{2,048\,a^5\,\ton{1 -\,e^2}^4}\times \\ \nonumber \\
&\times \grf{28\,e^2\,\ton{7 + 9\,\cos 2  I }\,\sin^3 I\,\cos 3 \omega + \ton{4 + 3\,e^2}\,\qua{2\,\sin I  +
7\,\ton{\sin 3  I  + 3\,\sin 5  I }}\,\cos \omega }, \\ \nonumber \\
\dot e_{.6} \nonumber &= \rp{105\,e\,\nk\,R_\mathrm{e}^6\,\sin^2 I}{8,192\,a^6\,\ton{1 -\,e^2}^5}\times\\ \nonumber\\
&\times \qua{5\,\ton{2 + e^2}\,\ton{35 + 60\,\cos 2  I  + 33\,\cos 4  I }\,\sin 2 \omega  + 12\,e^2\,\ton{9 + 11\,\cos 2  I }\,\sin^2 I\,\sin 4 \omega }, \\ \nonumber \\
\dot e_{.7} \nonumber &= \rp{21\,\nk\,R_\mathrm{e}^7}{524,288\,a^7\,\ton{1 -\,e^2}^6}\times \\ \nonumber\\
\nonumber &\times \grf{60\,e^2\,\sin^3 I\,\qua{-3\,\ton{8 + 3\,e^2}\,\ton{189 + 308\,\cos 2  I  + 143\,\cos 4  I }\,\cos 3 \omega  - \right.\right.\\ \nonumber \\
\nonumber &\left.\left. - 44\,e^2\,\ton{11 + 13\,\cos 2  I }\,\sin^2 I\,\cos 5\omega } - \right.\\ \nonumber \\
&\left. - 5\,\qua{8 + 5\,e^2\,\ton{4 + e^2}}\,\ton{25\,\sin I  + 81\,\sin 3  I  + 165\,\sin 5  I  + 429\,\sin 7  I }\,\cos\omega}, \\ \nonumber \\
\dot e_{.8} \nonumber &= -\rp{63\,e\,\nk\,R_\mathrm{e}^8\,\sin^2 I}{2,097,152\,a^8\,\ton{1 -\,e^2}^7}\times \\ \nonumber \\
\nonumber &\times \grf{35\,\ton{48 + 80\,e^2 + 15\,e^4}\,\ton{210 + 385\,\cos 2  I  + 286\,\cos 4  I  + 143\,\cos 6  I }\,\sin 2 \omega
+ \right.\\ \nonumber \\
\nonumber &\left. + 88\,e^2\,\sin^2 I \,\qua{7\,\ton{10 + 3\,e^2}\,\ton{99 + 156\,\cos 2  I  +65\,\cos 4  I }\,\sin 4 \omega  + \right.\right.\\ \nonumber \\
&\left.\left. + 26\,e^2\,\ton{13 + 15\,\cos 2  I }\,\sin^2 I\,\sin 6 \omega }}.
\end{align}
\subsection{The inclination}
\begin{align}
\dot I_{.2} &= 0, \\ \nonumber \\
\dot I_{.3} &= \rp{3\,e\,\nk\,R_\mathrm{e}^3\,\cos I\,\ton{3 + 5\,\cos 2  I }\,\cos \omega }{16\,a^3\,\ton{1 -\,e^2}^3}, \\ \nonumber \\
\dot I_{.4} &= \rp{15\,e^2\,\nk\,R_\mathrm{e}^4\,\cos I\,\ton{5 + 7\,\cos 2  I }\,\sin I\,\sin 2 \omega  }{64\,a^4\,\ton{1 -\,e^2}^4}, \\ \nonumber \\
\dot I_{.5} \nonumber &= -\rp{15\,e\,\nk\,R_\mathrm{e}^5\,\cot I}{2,048\,a^5\,\ton{1 -\,e^2}^5}\times \\ \nonumber \\
&\times \grf{28\,e^2\,\ton{7 + 9\,\cos 2  I }\,\sin^3 I\,\cos 3 \omega   + \ton{4 + 3\,e^2}\,\qua{2\,\sin I  +7\,\ton{\sin 3  I  + 3\,\sin 5  I }}\,\cos\omega}, \\ \nonumber \\
\dot I_{.6} \nonumber &= \rp{5\,\nk\,R_\mathrm{e}^6\,\cot I}{65,536\,a^6\,\ton{1 - \,e^2}^6}\times\\ \nonumber \\
\nonumber &\times \qua{-840\,e^2\,\ton{2 + e^2}\,\ton{35 + 60\,\cos 2  I  + 33\,\cos 4  I }\,\sin^2 I\,\sin 2 \omega  -\right.\\ \nonumber \\
&\left. - 2,016\,e^4\,\ton{9 + 11\,\cos 2  I }\,\sin^4 I\,\sin 4 \omega }, \\ \nonumber \\
\dot I_{.7} \nonumber &= -\rp{21\,e\,\nk\,R_\mathrm{e}^7\,\cot I}{524,288\,a^7\,\ton{1 - \,e^2}^7}\times\\ \nonumber \\
\nonumber &\times \grf{60\,e^2\,\sin^3 I\,\qua{-3\,\ton{8 + 3\,e^2}\,\ton{189 + 308\,\cos 2  I  + 143\,\cos 4  I}\,\cos 3 \omega -\right.\right.\\ \nonumber \\
\nonumber &\left.\left. - 44\,e^2\,\ton{11 + 13\,\cos 2  I }\,\sin^2 I\,\cos 5 \omega} -\right.\\ \nonumber \\
&\left. - 5\,\qua{8 + 5\,e^2\,\ton{4 + e^2}}\,\ton{25\,\sin I  + 81\,\sin 3  I  + 165\,\sin 5  I  + 429\,\sin 7  I}\,\cos\omega}, \\ \nonumber \\
\dot I_{.8} \nonumber &= \rp{63\,e^2\,\nk\,R_\mathrm{e}^8\,\sin I\,\cos I }{2,097,152\,a^8\,\ton{1 - \,e^2}^8}\times\\ \nonumber \\
\nonumber &\times \grf{35\,\ton{48 + 80\,e^2+ 15\,e^4}\,\ton{210 + 385\,\cos 2  I  +286\,\cos 4  I  + 143\,\cos 6  I }\,\sin 2 \omega  + \right.\\ \nonumber \\
\nonumber &\left. + 88\,e^2\,\sin^2 I\,\qua{7\,\ton{10 + 3\,e^2}\,\ton{99 + 156\,\cos 2  I  +65\,\cos 4  I}\,\sin 4 \omega  + \right.\right.\\ \nonumber \\
&\left.\left. + 26\,e^2\,\ton{13 + 15\,\cos 2  I }\,\sin^2 I \,\sin 6 \omega }}.
\end{align}
\subsection{The longitude of the ascending node}
\begin{align}
\dot\Omega_{.2} &= -\rp{3\,\nk\,R_\mathrm{e}^2\,\cos I  }{2\,a^2\,\ton{1 -\,e^2}^2}, \\ \nonumber \\
\dot\Omega_{.3} &= \rp{3\,e\,\nk\,R_\mathrm{e}^3\,\ton{\cos I  + 15\,\cos 3  I }\,\csc I\,\sin \omega  }{32\,a^3\,\ton{1 -\,e^2}^3}, \\ \nonumber \\
\dot\Omega_{.4} &= \rp{15\,\nk\,R_\mathrm{e}^4\,\qua{\ton{2 + 3\,e^2}\,\ton{9\,\cos I  + 7\,\cos 3  I } -2\,e^2\,\ton{5\,\cos I  + 7\,\cos 3  I }\,\cos 2 \omega } }{128\,a^4\,\ton{1 -\,e^2}^4}, \\ \nonumber \\
\dot\Omega_{.5} \nonumber &= -\rp{15\,e\,\nk\,R_\mathrm{e}^5}{2,048\,a^5\,\ton{1 -\,e^2}^5}\times \\ \nonumber \\
&\times \grf{\ton{4 + 3\,e^2} \qua{2\,\cos I  +21\,\ton{\cos 3  I  + 5\,\cos 5  I }} \csc I  \,\sin \omega  +7\,e^2\,\ton{2\,\sin 2  I  + 15\,\sin 4  I }\,\sin 3 \omega }, \\ \nonumber \\
\dot\Omega_{.6} \nonumber &= \rp{105\,\nk\,R_\mathrm{e}^6 }{16,384\,a^6\,\ton{1 -\,e^2}^6}\times \\ \nonumber \\
\nonumber &\times \qua{-\ton{8 + 40\,e^2+ 15\,e^4}\,\ton{50\,\cos I  + 45\,\cos 3  I  +33\,\cos 5  I } +\right.\\ \nonumber \\
\nonumber &\left.+ 5\,e^2\,\ton{2 + e^2}\,\ton{70\,\cos I  + 87\,\cos 3  I  +99\,\cos 5  I }\,\cos 2 \omega  +\right.\\ \nonumber \\
&\left. + 6\,e^4\,\ton{47\,\cos I  + 33\,\cos 3  I }\,\sin^2 I\,\cos 4 \omega}, \\ \nonumber \\
\dot\Omega_{.7} \nonumber &= -\rp{21\,e\,\nk\,R_\mathrm{e}^7\,\csc I}{524,288\,a^7\,\ton{1 - \,e^2}^7}\times \\ \nonumber \\
\nonumber &\times\grf{-5\,\qua{8 + 5\,e^2\,\ton{4 + e^2}}\,\ton{25\,\cos I  + 243\,\cos 3  I  + 825\,\cos 5  I  + 3,003\,\cos 7  I }\,\sin \omega  -\right.\\ \nonumber \\
\nonumber &\left. - 30\,e^2\,\ton{8 + 3\,e^2}\,\ton{1,442\,\cos I  + 1,397\,\cos 3  I  + 1,001\,\cos 5  I }\,\sin^2 I \,\sin 3 \omega  -\right.\\ \nonumber \\
&\left. - 264\,e^4\,\ton{149\,\cos I  + 91\,\cos 3  I }\,\sin^4 I \,\sin 5 \omega }, \\ \nonumber \\
\dot\Omega_{.8} \nonumber &= \rp{63\,\nk\,R_\mathrm{e}^8 }{2,097,152\,a^8\,\ton{1 -\,e^2}^8}\times \\ \nonumber \\
\nonumber &\times\ton{5\,\grf{16 + 7\,e^2\,\qua{24 + 5\,e^2\,\ton{6 + e^2}}}\,\qua{1,225\,\cos I  + 11\,\ton{105\,\cos 3  I  + 91\,\cos 5  I  + 65\,\cos 7  I }} - \right.\\ \nonumber \\
\nonumber &\left. - 70\,e^2\,\ton{48 + 80\,e^2+ 15\,e^4}\,\grf{105\,\cos I  +11 \qua{11\,\cos 3  I  + 13\,\ton{\cos 5  I  + \cos 7  I }}}\,\cos 2 \omega  -\right.\\ \nonumber \\
\nonumber &\left. - 616\,e^4\,\ton{10 + 3\,e^2}\,\ton{138\,\cos I  + 117\,\cos 3  I  +65\,\cos 5  I }\,\sin^2 I\,\cos 4 \omega - \right.\\ \nonumber \\
&\left. - 9,152\,e^6\,\cos I \,\ton{2 + 5\,\cos 2  I }\,\sin^4 I\,\cos 6 \omega }.
\end{align}
\subsection{The argument of perigee}
\begin{align}
\dot\omega_{.2} &= \rp{3\,\nk\,R_\mathrm{e}^2\,\ton{3 + 5\,\cos 2  I } }{8\,a^2\,\ton{1 -\,e^2}^2}, \\ \nonumber \\
\dot\omega_{.3} &= -\rp{3\,\nk\,R_\mathrm{e}^3}{64\,a^3\,e\,\ton{1 -\,e^2}^3} \qua{-1 - 3\,e^2\,- 4\,\cos 2  I  +5\,\ton{1 + 7\,e^2}\,\cos 4  I }\,\csc I  \,\sin \omega  , \\ \nonumber \\
\dot\omega_{.4} \nonumber &= \rp{15\,\nk\,R_\mathrm{e}^4}{1,024\,a^4\,\ton{1 -\,e^2}^4}\times\\ \nonumber \\
\nonumber &\times \grf{-27\,\ton{4 + 5\,e^2} + 4\,\cos 2  I \,\qua{-52 - 63\,e^2+ 2\,\ton{-2 + 7\,e^2}\,\cos 2 \omega } + 2\,\ton{-6 + 5\,e^2}\,\cos 2 \omega  +\right.\\ \nonumber \\
&\left. + 7\,\cos 4  I \,\qua{-28 - 27\,e^2+ 2\,\ton{2 + 9\,e^2}\,\cos 2 \omega }}, \\ \nonumber \\
\dot\omega_{.5} \nonumber &= -\rp{15\,\nk\,R_\mathrm{e}^5\,\sin I}{4,096\,a^5\,e\,\ton{1 - e^2}^5}\times\\ \nonumber \\
\nonumber &\times \grf{\qua{8 + 74\,e^2+ 30\,e^4\,+ \ton{20 + 113\,e^2+ 21\,e^4}\,\cos 2  I  + 14\,\ton{4 + 5\,e^2\,- 9\,e^4}\,\cos 4  I  -\right.\right.\\ \nonumber \\
\nonumber &\left.\left. - 21\,\ton{4 + 61\,e^2+ 33\,e^4}\,\cos 6  I }\,\csc^2 I\,\sin \omega  -\right.\\ \nonumber \\
&\left. - 14\,e^2\,\qua{-5 + 7\,e^2+4\,\ton{-1 + 6\,e^2}\,\cos 2  I  + \ton{9 + 33\,e^2}\,\cos 4  I }\,\sin 3 \omega }, \\ \nonumber \\
\dot\omega_{.6} \nonumber &= \rp{105\,\nk\,R_\mathrm{e}^6}{65,536\,a^6\,\ton{1 -\,e^2}^6}\times \\ \nonumber \\
\nonumber &\times \ton{5\,\grf{\ton{472 + 1,940\,e^2+ 675\,e^4}\,\cos 2  I  + \right.\right.\\ \nonumber \\
\nonumber &\left.\left. + 3\,e^2\,\qua{2\,\ton{292 + 99\,e^2}\,\cos 4  I  +11\,\ton{44 + 13\,e^2}\,\cos 6  I }} -\right.\\ \nonumber \\
\nonumber &\left. - 5\,\qua{10\,e^2\,\ton{6 + 7\,e^2} + \ton{-68 + 254\,e^2+ 195\,e^4}\,\cos 2  I  +6\,\ton{-4 + 102\,e^2+ 55\,e^4}\,\cos 4I  +\right.\right.\\ \nonumber \\
\nonumber &\left.\left. + 33\,\ton{4 + 34\,e^2+ 13\,e^4}\,\cos 6  I }\,\cos 2 \omega  +\right.\\ \nonumber \\
\nonumber &\left. + 2\,\qua{1,128\,\cos 4  I  + 1,188\,\cos 6  I  +25\,\ton{24 + 100\,e^2+ 35\,e^4\,+ 4\,\cos 2 \omega }} -\right.\\ \nonumber \\
&\left. - 6\,e^2\,\qua{-28 + 45\,e^2+ 4\,\ton{-4 + 33\,e^2}\,\cos 2  I  +11\,\ton{4 + 13\,e^2}\,\cos 4  I }\,\sin^2 I\,\cos 4 \omega }, \\ \nonumber \\
\dot\omega_{.7} \nonumber &= -\rp{21\,\nk\,R_\mathrm{e}^7 }{524,288\,e\,a^7\,\ton{1 - \,e^2}^7}\times \\ \nonumber \\
\nonumber &\times \ton{-5\,\ton{8 + 156\,e^2+ 225\,e^4\,+ 40\,e^6}\,\ton{25\,\sin I + 81\,\sin 3  I  + 165\,\sin 5  I  +429\,\sin 7  I }\,\sin \omega  -\right.\\ \nonumber \\
\nonumber &\left. - 60\,e^2\,\ton{24 + 95\,e^2+ 24\,e^4}\,\ton{189 + 308\,\cos 2  I  + 143\,\cos 4  I }\,\sin^3 I \,\sin 3 \omega  +\right.\\ \nonumber \\
\nonumber &\left. + 10\,e^2\,\cos I \,\grf{\qua{8 + 5\,e^2\,\ton{4 + e^2}}\,\ton{-1,198 + 2,421\,\cos 2  I  -2,178\,\cos 4  I  + \right.\right.\right.\\ \nonumber \\
\nonumber &\left.\left.\left. + 3,003\,\cos 6  I }\cot I \,\sin \omega  +\right.\right.\\ \nonumber \\
\nonumber &\left.\left. + 3\,e^2\,\ton{8 + 3\,e^2}\,\ton{523 + 396\,\cos 2  I  +1,001\,\cos 4  I }\,\sin 2  I \,\sin 3 \omega } +\right.\\ \nonumber \\
\nonumber &\left. + 528\,e^6\,\cos^2 I \,\ton{29 + 91\,\cos 2  I }\,\sin^3 I \,\sin 5 \omega - \right.\\ \nonumber \\
&\left. - 528\,e^4\,\ton{5 + 8\,e^2}\,\ton{11 + 13\,\cos 2  I }\,\sin^5 I \,\sin 5 \omega }, \\ \nonumber \\
\dot\omega_{.8} \nonumber &= \rp{63\,\nk\,R_\mathrm{e}^8}{33,554,432\,a^8\,\ton{1 -\,e^2}^8}\times \\ \nonumber \\
\nonumber &\times \ton{5\,\grf{-1,225\,\qua{192 + 35\,e^2\,\ton{48 + 56\,e^2+ 9\,e^4}} -\right.\right.\\ \nonumber \\
\nonumber &\left.\left. - 280\,\qua{1,664 + 7\,e^2\,\ton{2,064 + 2,400\,e^2+ 385\,e^4}}\,\cos 2  I  - \right.\right.\\ \nonumber \\
\nonumber &\left.\left. - 308\,\qua{1,472 + 7\,e^2\,\ton{1,776 + 2,040\,e^2+ 325\,e^4}}\,\cos 4  I  -\right.\right.\\ \nonumber \\
\nonumber &\left.\left. - 3,432\,\qua{128 + 7\,e^2\,\ton{144 + 160\,e^2+ 25\,e^4}}\,\cos 6  I  -\right.\right.\\ \nonumber \\
\nonumber &\left.\left. - 715\,\qua{704 + 7\,e^2\,\ton{624 + 600\,e^2+ 85\,e^4}}\,\cos 8  I } + \right.\\ \nonumber \\
\nonumber &\left. + 70\,\qua{35\,\ton{-96 + 208\,e^2+ 950\,e^4\,+ 225\,e^6} + \right.\right.\\ \nonumber \\
\nonumber &\left.\left. + 16\,\ton{-384 + 1,648\,e^2+ 5,160\,e^4\,+ 1,155\,e^6}\,\cos 2  I  + \right.\right. \\ \nonumber \\
\nonumber &\left.\left. + 44\,\ton{-96 + 1,360\,e^2+ 2,870\,e^4\,+ 585\,e^6}\,\cos 4  I  + \right.\right.\\ \nonumber \\
\nonumber &\left.\left. + 2,288\,e^2\,\ton{48 + 80\,e^2+ 15\,e^4}\,\cos 6  I  + \right.\right. \\ \nonumber \\
\nonumber &\left.\left. + 143\,\ton{96 + 1,328\,e^2+ 1,610\,e^4\,+ 255\,e^6}\,\cos 8  I }\,\cos 2 \omega  + \right.\\ \nonumber \\
\nonumber &\left. + 616\,e^2\,\qua{6\,\ton{-280 + 944\,e^2+ 363\,e^4} + \ton{-1,960 + 14,128\,e^2+4,797\,e^4}\,\cos 2  I  +\right.\right.\\ \nonumber \\
\nonumber &\left.\left. + 26\,\ton{40 + 688\,e^2+ 195\,e^4}\,\cos 4  I  +65\,\ton{40 + 208\,e^2+ 51\,e^4}\,\cos 6  I }\,\sin^2 I\,\cos 4 \omega + \right.\\ \nonumber \\
&\left. +  4,576\,e^4\,\qua{-22 + 39\,e^2+ 4\,\ton{-2 + 25\,e^2}\,\cos 2  I  +5\,\ton{6 + 17\,e^2}\,\cos 4  I }\,\sin^4 I\,\cos 6 \omega }.
\end{align}
\subsection{The mean anomaly at epoch}
\begin{align}
\dot\eta_{.2} &= \rp{3\,\nk\,R_\mathrm{e}^2\,\ton{1 + 3\,\cos 2  I }}{8\,a^2\,\ton{1 -\,e^2}^{3/2}}, \\ \nonumber \\
\dot\eta_{.3} &= \rp{3\,\nk\,R_\mathrm{e}^3}{16\,e\,a^3\,\ton{1 -\,e^2}^{5/2}}\qua{\ton{-1 + 4\,e^2}\,\ton{3 + 5\,\cos 2  I }\,\sin I \,\sin \omega }, \\ \nonumber \\
\dot\eta_{.4} \nonumber &= -\rp{15\,\nk\,R_\mathrm{e}^4}{1,024\,a^4\,\ton{1 -\,e^2}^{7/2}}\times\\ \nonumber \\
&\times\qua{3\,e^2\,\ton{9 + 20\,\cos 2  I  + 35\,\cos 4  I } +8\,\ton{-2 + 5\,e^2}\,\ton{5 + 7\,\cos 2  I }\,\sin^2 I\,\cos 2 \omega }, \\ \nonumber \\
\dot\eta_{.5} \nonumber &= \rp{15\,\nk\,R_\mathrm{e}^5 }{2,048\,e\,a^5\,\ton{1 -\,e^2}^{9/2}}\times \\ \nonumber \\
\nonumber &\times \grf{-\ton{-4 + 7\,e^2+ 18\,e^4} \qua{2\,\sin I  +7\,\ton{\sin 3  I  + 3\,\sin 5  I }}\,\sin \omega  -\right.\\ \nonumber \\
&\left. - 28\,e^2\,\ton{-1 + 2\,e^2}\,\ton{7 + 9\,\cos 2  I }\,\sin^3 I \,\sin 3 \omega }, \\ \nonumber \\
\dot\eta_{.6} \nonumber &= -\rp{35 \,\nk\,R_\mathrm{e}^6}{65,536\,a^6\,\ton{1 -\,e^2}^{11/2}}\times\\ \nonumber \\
\nonumber &\times \qua{-\ton{-8 + 20\,e^2+ 15\,e^4}\,\ton{50 + 105\,\cos 2  I  +126\,\cos 4  I  + 231\,\cos 6  I } -\right.\\ \nonumber \\
\nonumber &\left. - 60\,\ton{-4 + 6\,e^2+ 7\,e^4}\,\ton{35 + 60\,\cos 2  I  + 33\,\cos 4  I }\,\sin^2 I\,\cos 2 \omega   -\right.\\ \nonumber \\
&\left.- 72\,e^2\,\ton{-4 + 7\,e^2}\,\ton{9 + 11\,\cos 2  I }\,\sin^4 I\,\cos 4 \omega }, \\ \nonumber \\
\dot\eta_{.7} \nonumber &= \rp{21\,\nk\,R_\mathrm{e}^7}{524,288\,e\,a^7\,\ton{1 -\,e^2}^{13/2}}\times \\ \nonumber \\
\nonumber &\times \qua{5\,\ton{-8 - 28\,e^2+ 95\,e^4\,+ 40\,e^6}\,\ton{25\,\sin I  +81\,\sin 3  I  + 165\,\sin 5  I  +429\,\sin 7  I }\,\sin \omega  +\right.\\ \nonumber \\
\nonumber &\left. + 180\,e^2\,\ton{-8 + 11\,e^2+ 8\,e^4}\,\ton{189 + 308\,\cos 2  I  +143\,\cos 4  I }\,\sin^3 I \,\sin 3 \omega  +\right.\\ \nonumber \\
&\left. + 528\,e^4\,\ton{-5 + 8\,e^2}\,\ton{11 + 13\,\cos 2  I }\,\sin^5 I \,\sin 5 \omega }, \\ \nonumber \\
\dot\eta_{.8} \nonumber &= -\rp{63\,\nk\,R_\mathrm{e}^8}{33,554,432\,a^8\,\ton{1 -\,e^2}^{15/2}}\times \\ \nonumber \\
\nonumber &\times \grf{5 \qua{-32 + 35\,e^4\,\ton{4 + e^2}}\,\ton{1,225 + 2,520\,\cos 2  I  +2,772\,\cos 4  I  + 3,432\,\cos 6  I  + \right.\right.\\ \nonumber \\
\nonumber &\left.\left. + 6,435\,\cos 8  I } +280\,\ton{-96 - 80\,e^2+ 470\,e^4\,+ 135\,e^6}\,\ton{210 + 385\,\cos 2  I  +286\,\cos 4  I  + \right.\right.\\ \nonumber \\
\nonumber &\left.\left. + 143\,\cos 6  I }\,\sin^2 I\,\cos 2 \omega  +\right.\\ \nonumber \\
\nonumber &\left. + 2,464\,e^2\,\ton{-40 + 52\,e^2+ 27\,e^4}\,\ton{99 + 156\,\cos 2  I  +65\,\cos 4  I }\,\sin^4 I\,\cos 4 \omega  +\right.\\ \nonumber \\
&\left. + 18,304\,e^4\,\ton{-2 + 3\,e^2}\,\ton{13 + 15\,\cos 2  I }\,\sin^6 I\,\cos 6 \omega }.
\end{align}
\section{The atmospheric drag}\lb{appendrag}
\renewcommand{\theequation}{C\arabic{equation}}
\setcounter{equation}{0}
The neutral and charged atmospheric drag is potentially a major source of systematic uncertainty since it induces long-term effects on all the
satellite's orbital elements which, for a frozen perigee configuration, may look like secular trends on which the time variability of the atmospheric density is superimposed.

For the sake of simplicity, we will model HERO as a cannonball, passive satellite with the same physical properties of the existing LARES satellite  in order to make some quantitative estimates of the disturbing impact of the atmospheric drag on the pN effects of interest.  Thus, its perturbing acceleration is customarily modeled as
\eqi
{\bds A}_\textrm{D} = -\rp{1}{2}\,C_\textrm{D}\,\Sigma\,\rho\,V\,\bds V,\lb{adrag}
\eqf
where $C_\textrm{D},~\Sigma,~\rho,~\bds V$ are the dimensionless drag coefficient of the spacecraft, its area-to-mass ratio, the atmospheric density at its height, and its velocity with respect to the atmosphere, respectively. By assuming that the atmosphere co-rotates with the Earth, $\bds V$ can be posed equal to
\eqi
\bds V = \bds{\textrm{v}}-\bds\Psi\cross\bds r,\lb{roto}
\eqf
where $\bds\Psi$ is the Earth's angular velocity. In fact, a decrease of the co-rotation with the height is expected. \citet{2010EJPh...31.1013M} modeled it in two scenarios involving  a constant and non-constant viscosity. In the first case, the second term in \rfr{roto} must be rescaled by $\ton{R_\textrm{e}/r}^3$.
In general, the atmospheric density may not be considered as constant throughout a highly eccentric orbit covering a wide range of geocentric distances, as in our case; thus, we will model it as
\eqi
\rho(r) = \rho_0\exp\qua{-\rp{\ton{r-r_0}}{\lambda}},\lb{densi}
\eqf
where $\rho_0$ is the atmospheric density at some reference distance $r_0$, while $\lambda$ is the characteristic scale length. By assuming
\eqi
r_0=r_\textrm{min}=a\left(1-e\right),
\eqf
if the atmospheric density is known at the perigee and apogee heights, $\lambda$ can be determined as
\eqi
\lambda = -\rp{2ae}{\ln\ton{\rp{\rho_\textrm{min}}{\rho_\textrm{max}}}},\lb{Lam}
\eqf
where
\begin{align}
\rho_\textrm{min}&=\rho(r_\textrm{max}),\\ \nonumber \\
\rho_\textrm{max} & =\rho(r_\textrm{min}).
\end{align}

In the case of the orbital configuration of Table\,\ref{tavola1}, the perigee height is $h_\mathrm{min}=1,046.86\,\mathrm{km}$; we will determine the corresponding neutral atmospheric density $\rho_0=\rho_\mathrm{max}$ as follows.
The neutral atmospheric density at the LARES height, which is $h_\mathrm{LR} = 1,450\,\mathrm{km}$, amounts to $\rho_\mathrm{LR} =5.644\times 10^{-16}\,\mathrm{kg\,m}^{-3}$ \citep{2017AcAau.140..469P}. According to \citet{2015JPhCS.641a2026B}, the neutral atmospheric density at $h_{700}=700\,\mathrm{km}$ is  $6.9\times 10^{-14}\,\mathrm{kg\,m}^{-3}$ (TD-88 model), or $1.11\times 10^{-14}\,\mathrm{kg\,m}^{-3}$ (NASA model). Thus, it is possible to calculate the characteristic scale length $\lambda$ valid for the range $700\,\mathrm{km}< h< 1,450\,\mathrm{km}$ as
\eqi
\lambda_\mathrm{700/LR} = -\rp{\ton{h_{700} - h_\mathrm{LR}}}{\ln\ton{\rp{\rho_\mathrm{LR}}{\rho_{700}}}} = 154.422\,\mathrm{km}-249.139 \,\mathrm{km},\lb{L700}
\eqf
depending on the value adopted for $\rho$ at $700\,\mathrm{km}$.
Since for the orbital configuration of HERO of Table\,\ref{tavola1} it is $h_{700}< h_\mathrm{min} < h_\mathrm{LR}$, one can determine $\rho_\mathrm{max}$ by using just \rfr{L700}  in
\eqi
\rho_\mathrm{max} = \rho_\mathrm{LR}\,\exp\qua{-\rp{\ton{h_\mathrm{LR}-h_\mathrm{min}}}{\lambda_\mathrm{700/LR}}} =\ton{7.3-2.8}\times 10^{-15}\,\mathrm{kg\,m}^{-3},\lb{rhomax}
\eqf
depending on the value of $\lambda_\mathrm{700/LR}$ adopted.
As far as $\rho_\mathrm{min}$ is concerned, since $h_\mathrm{max}=13,169.9\,\mathrm{km}$ is much larger than the height of, say, the LAGEOS satellite ($h_\mathrm{L}=5,891.87\,\mathrm{km}$), for which it is $\rho_\mathrm{L}=6.579\times 10^{-18}\,\mathrm{kg\,m}^{-3}$ \citep{2015CQGra..32o5012L},
it does not seem unreasonable to assume
\eqi
\rho_\mathrm{min}\simeq 0.001\,\rho_\mathrm{L},\lb{rhomin}
\eqf
or so.
Thus, \rfr{Lam}, applied to $\rho_\mathrm{max}$, given by \rfr{rhomax}, and $\rho_\mathrm{min}$, given by \rfr{rhomin}, allows to infer the value for $\lambda$ which must be used in the calculation of the drag effect for HERO. It is
\eqi
\lambda = 872.87\,\mathrm{km}-938.49\,\mathrm{km},
\eqf
depending on the value of $\rho_\mathrm{max}$ adopted.
As far as the more eccentric orbital configuration of Table\,\ref{tavola3} is concerned, by adopting for $\rho_0$ the two possible values of $\rho_{700}$ and, say, $\rho_\mathrm{min}=0.0001\,\rho_\mathrm{L}$, it turns out that
\eqi
\lambda = 3,463.23\,\mathrm{km}-3,843.48\,\mathrm{km}.
\eqf

Actually, even the density at a given height may not be regarded as truly constant because of a variety of geophysical phenomena characterized by quite different time scales. Anyway, in order to have an order-of-magnitude evaluation of the perturbing action of \rfr{adrag} on the motion of HERO, we calculate the averaged rates of change of its Keplerian orbital elements by keeping $\rho_0=\rho_\mathrm{max}$ in \rfr{densi}
fixed during one orbital period $P_\textrm{b}$; given its short duration, at least in the case of Table\,\ref{tavola1}, it is not an unreasonable assumption. The large value of the eccentricity makes most of the existing results in the literature unsuitable to the present case; moreover, an exact analytical calculation without recurring to any approximation in both $e$ and $\nu\doteq \Psi/\nk$ is difficult.
Thus, we follow two complementary approaches. In one of them, we, first, plot in Figure\,\ref{fig1} and Figure\,\ref{fig3} the analytical expressions of the ratios of the right-hand-sides of the Gauss perturbing equations for the rates of change of the Keplerian orbital elements, evaluated onto the unperturbed Keplerian ellipse, to $\Pb$
as  functions of the true anomaly $f$.
In the most general case, by assuming the co-rotation of the atmosphere, they are
\begin{align}
\rp{\nk}{2\,\uppi}\dert{a}{f} \lb{uno}& = -\rp{ C_\textrm{D}\,\Sigma\,\rho\,a\,\nk\,\sqrt{1-e^2}\,\mathcal{V}\,\qua{1 + 2\,e\,\cos f + e^2 - \nu\,\ton{1-e^2}^{3/2}\,\cos I} }{ 2\,\uppi\,\ton{1 + e\,\cos f}^2 }, \\ \nonumber \\
\rp{\nk}{2\,\uppi}\dert{e}{f} \nonumber & = -\rp{ C_\textrm{D}\,\Sigma\,\rho\,\nk\,\ton{1-e^2}^{3/2}\,\mathcal{V} }{ 8\,\uppi\,\ton{1 + e\,\cos f}^4 }\grf{
4\,\ton{e+\cos f}\,\ton{1 + e\,\cos f}^2 -\right.\\ \nonumber \\
&\left.  - \nu\,\rp{\ton{1-e^2}^{3/2}\,\qua{4\,\cos f + e\,\ton{3 + \cos 2 f} } }{\sqrt{5}}
}\lb{due}, \\ \nonumber \\
\rp{\nk}{2\,\uppi}\dert{I}{f} \lb{tre}& = -\rp{ C_\textrm{D}\,\Sigma\,\rho\,\nk\,\nu\,\ton{1-e^2}^3\,\mathcal{V}\,\sin I\,\cos^2 u}{ 4\,\uppi\,\ton{1 + e\,\cos f}^4 }, \\ \nonumber \\
\rp{\nk}{2\,\uppi}\dert{\Omega}{f} \lb{quattro}& = -\rp{ C_\textrm{D}\,\Sigma\,\rho\,\nk\,\nu\,\ton{1-e^2}^3\,\mathcal{V}\,\sin 2 u}{ 8\,\uppi\,\ton{1 + e\,\cos f}^4 }, \\ \nonumber \\
\rp{\nk}{2\,\uppi}\dert{\omega}{f} \nonumber & = \rp{ C_\textrm{D}\,\Sigma\,\rho\,\nk\,\ton{1-e^2}\,\mathcal{V}}{ 8\,\uppi\,e\,\ton{1 + e\,\cos f}^4 }\,\grf{ -4\,\sqrt{1 - e^2}
\ton{1 + e\,\cos f}^2\,\sin f + \right.\\ \nonumber \\
&\left. +2\,\nu\,\ton{1 - e^2}^2\,\cos I\,\qua{2\,\sin f + e\,\cos\omega\,\sin\ton{2f+\omega}} }\lb{cinque},\\ \nonumber \\
\rp{\nk}{2\,\uppi}\dert{\eta}{f} \nonumber \lb{sei}& = \rp{ C_\textrm{D}\,\Sigma\,\rho\,\nk\,\ton{1-e^2}^2\,\mathcal{V}\,\sin f}{ 4\,\uppi\,e\,\ton{1 + e\,\cos f}^4 }\,\qua{2 + 3\,e^2 + 2\,e\,\ton{2 + e^2}\,\cos f + e^2\,\cos 2 f - \right.\\ \nonumber \\
&\left. - \nu\,\ton{1 - e^2}^{3/2}\,\ton{2 + e\,\cos f}\,\cos I},
\end{align}
where
\begin{align}
{\mathcal{V}}^2 \lb{Vu}& =1 - \nu\,\rp{2\,\ton{1 - e^2}^{3/2}\,\cos I}{1 + e^2 + 2\,e\,\cos f} +\nu^2\,\rp{\ton{1-e^2}^3\,\ton{3 + \cos 2 I +2\,\sin^2 I\,\cos 2u } }{4\,\ton{1 + e\,\cos f}^2\,\ton{1 + e^2 + 2\,e\,\cos f} },
\end{align}
and
\eqi u\doteq \omega+ f
\eqf
is the argument of latitude.
It is worthwhile noticing that, in general,
\eqi
\left|{\mathcal{V}}^2-1\right|\nless 1,
\eqf
being even possible that
\eqi
\left|{\mathcal{V}}^2-1\right|\gtrsim 1
\eqf  for some values of $f$, thus preventing from expanding it in powers of $\nu$.
Then,  we numerically calculate the areas under the resulting curves, i.e. we  numerically integrate the averaged rates of change of the orbital elements for the given orbital configuration: see Table\,\ref{tavola2ter} and Table\,\ref{tavola4ter}. In the second approach, we numerically integrate the equations of motion of the satellite in rectangular Cartesian coordinates, referred to a geocentric equatorial coordinate system, with and without \rfr{adrag} over 1 yr; both the runs share the same initial conditions. Then, we subtract the resulting  time series of the orbital elements in order to single out their shifts due to the disturbing acceleration. Finally, we perform a linear fit of them, and look at their slopes; we plot the fitted trends as functions of time $t$ in Figure\,\ref{fig2} and Figure\,\ref{fig4}.  Both the methods reciprocally agree well, as shown by Figures\,\ref{fig1}\,to\,\ref{fig2} and Figures\,\ref{fig3}\,to\,\ref{fig4} for the neutral drag.
A slight reduction turns out to occur if the decrease of the atmospheric co-rotation with height is taken into account as modeled by \citet{2010EJPh...31.1013M} by assuming the simpler case of a constant viscosity.
We tested our approach by checking that it was able to reproduce the observed features of the semimajor axis decay of LARES recently determined in \citet{2017AcAau.140..469P}.
However, it must be stressed that such findings should be deemed just as indicative of the limitations of the scenario considered if a passive spacecraft were to be adopted. Indeed, they were computed preliminarily by assuming the same physical properties of the existing LARES satellite which, in principle, could well be superseded by a new, specifically manufactured spacecraft able to reduce both the drag coefficient $C_\textrm{D}$ and the area-to-mass ratio $\Sigma$. Moreover, also the actual temporal variability of the atmospheric density over timescales larger than  the satellite's orbital period $\Pb$ should be taken into account, especially if data will be collected during temporal intervals several years long.
To this aim, it is important to note that an inspection of \rfrs{uno}{sei} shows that no other sources of long-term modulation are present.
Indeed, the circulating node $\Omega$ does not enter them, contrary to the perigee $\omega$ which, however, is held fixed by the adopted value of the inclination $I$.
\end{appendices}
\bibliography{MS_binary_pulsar_bib,Gclockbib,semimabib,PXbib}{}

\end{document}